\documentclass[pra,reprint,amsmath,amssymb,aps,superscriptaddress]{revtex4-1}

\usepackage{lipsum}
\usepackage{graphicx}% Include figure files
\usepackage{dcolumn}% Align table columns on decimal point
\usepackage{bm}% bold math

\usepackage{xargs}
\usepackage[pdftex,divipsnames]{xcolor}
\usepackage[colorinlistoftodos,prependcaption,textsize=tiny]{todonotes}
\newcommandx{\tdunsure}[2][1=]{\todo[linecolor=red,backgroundcolor=red!25,bordercolor=red,#1]{#2}}
\newcommandx{\tdcomment}[2][1=]{\todo[linecolor=blue,backgroundcolor=blue!25,bordercolor=blue,#1]{#2}}
\newcommandx{\tdadd}[2][1=]{\todo[linecolor=yellow,backgroundcolor=yellow!25,bordercolor=yellow,#1]{#2}}

\usepackage{amsfonts}

\usepackage[caption=false]{subfig}

\usepackage[hidelinks]{hyperref}

\usepackage{cleveref}
\crefname{section}{\S}{\SS}
\crefformat{figure}{#2Fig. #1#3}
\crefformat{table}{#2Table #1#3}
\crefformat{equation}{#2Eq. (#1)#3}

\newcommand{\figwidth}{0.7}

\begin{document}

\title{Dynamics and Stability of an Optically Levitated Mirror}

\author{Ruvi Lecamwasam}
\email[]{ruvi.lecamwasam@anu.edu.au}

\author{Alistair Graham}
\author{Jinyong Ma}
\author{Kabilan Sripathy}
\author{Giovanni Guccione}
\author{Jiayi Qin}
\author{Geoff Campbell}
\author{Ben Buchler}
\author{Joseph Hope}
\author{Ping Koy Lam}

\affiliation{Centre for Quantum Computation and Communication Technology, Department of Quantum Science, Research School of Physics and Engineering, The Australian National University, Canberra ACT 2601 Australia}
\date{\today}% It is always \today, today,

%}]}

\begin{abstract}
We analyse the dynamics of a one-dimensional vertical Fabry-P\'erot cavity, where the upper mirror levitates due to intra-cavity radiation pressure force. A perturbative approach is used based around separation of timescales, which allows us to calculate the physical quantities of interest. Due to the dynamics of the cavity field, we find that the upper mirror's motion will always be unstable for levitation performed using only a single laser. Stability can be achieved for two lasers, where one provides the trapping potential and the other a damping effect, and we locate and characterise all parameter regimes where this can occur. Finally we analyse photothermal effects due to heating of the mirror substrate. We show that this can stabilise the system, even with only a single input laser, if it acts to increase the optical path length of the cavity. This work serves as a foundation for understanding how levitated optical cavity schemes can be used as stable metrological platforms.
\end{abstract}

\maketitle

%{[{ Main body
\section{Introduction}\label{secIntroduction}%{[{

Optomechanics explores the interaction between electromagnetic radiation and mechanical motion. The canonical example is a Fabry-P\'erot cavity, one end of which oscillates on a spring \cite{Aspelmeyer2014}. This interaction between different physical systems opens up many applications. We can perform precision metrology by coupling the mechanical motion to some force of interest and reading out its position via the cavity field \cite{Pontin2014}, and thermal motion of the oscillator can be cooled using optical techniques \cite{Groblacher2009}. A quantum state of light in the cavity could generate a macroscopic quantum superposition in the mechanical oscillator \cite{Mancini1997,Fiore2011}, providing a platform for tests of quantum decoherence \cite{Bose1999} and models of semiclassical quantum gravity \cite{Yang2013,Gan2016,Grossardt2016}. Alternatively, the interaction can lead to the generation of squeezed light for quantum information \cite{Marino2010,Safavi2013}.

The main source of noise and decoherence in optomechanical systems is generally thermal effects from the environment, which will couple into the system via the mechanical oscillator and can be significant even at cryogenic temperatures \cite{Hao2003}. One solution is to replace the mechanical spring with optical trapping and levitation of one of the cavity mirrors \cite{Libbrecht2004,Arvanitaki2013,Kiesel2013,Romero2011,Singh2010,Corbitt2007}, which can also lead to optical springs with much higher $Q$ factors than is possible mechanically \cite{Ni2012}. In particular, we will consider a vertically oriented Fabry-P\'erot cavity where the upper mirror levitates on the intracavity field, and is thus decoupled from the environment.

Such a system has been proposed and analysed \cite{Guccione2013,Michimura2017,Gan2016,Ho2019}, however a detailed investigation of the optical spring has not yet been performed. While the levitating mirror shares many characteristics with conventional optomechanical systems, the lack of a reference mechanical oscillator means that the spring is entirely optical in nature. We will show that this introduces some key differences in behaviour.

In this paper we perform a theoretical investigation of the optical spring of a one-dimensional levitating system, making use of a perturbative and separation-of-timescales approach. While our analysis will be fully classical, such a treatment is a necessary precursor to both a fully quantum investigation and experimental realisation. We will be concerned only with optically-induced stability, with the assumption that geometric stability has already been provided. This could be done via a field gradient as in optical tweezers \cite{Ashkin1970,Verdeny2011}, a multi-beam configuration \cite{Guccione2013}, or other means such as electromagnetic confinement.

In \cref{sec:MainAnalysis} we introduce a perturbative analysis based on separation of the optical and mechanical timescales. Key dimensionless parameters are identified in \cref{sec:MainNondimensionalisation}, and an adiabatic approximation in \cref{sec:MainAdiabatic} allows us to identify an effective potential in which the levitating mirror moves. In \cref{sec:MainDynamics} we consider the first-order correction to this which introduces optical anti-damping to the system, and in \cref{sec:MainVisualisation} we show how the previous analysis leads to an intuitive visualisation of the evolution of our dynamical variables. We then apply this formalism in \cref{sec:Application}, analysing passive two-laser optical cooling in \cref{sec:ApplicationTwoLaser} and photothermal expansion in \cref{sec:ApplicationPhotothermal}.
 
Quantities defined in the paper are summarised in \cref{tab:Parameters} and \cref{tab:NonDimParameters}. Code for computer simulations, their results, and all figures, may be downloaded from ref. \cite{supplementaryGithub}.

%}]}
\section{Analysis of the ideal system}\label{sec:MainAnalysis}%{[{
We first consider an idealised version of our system. This can be very well characterised, and our findings and intuition can be translated to when we add in experimental considerations such as additional lasers and photothermal expansion.

\begin{figure}
  \includegraphics[width=0.6\columnwidth]{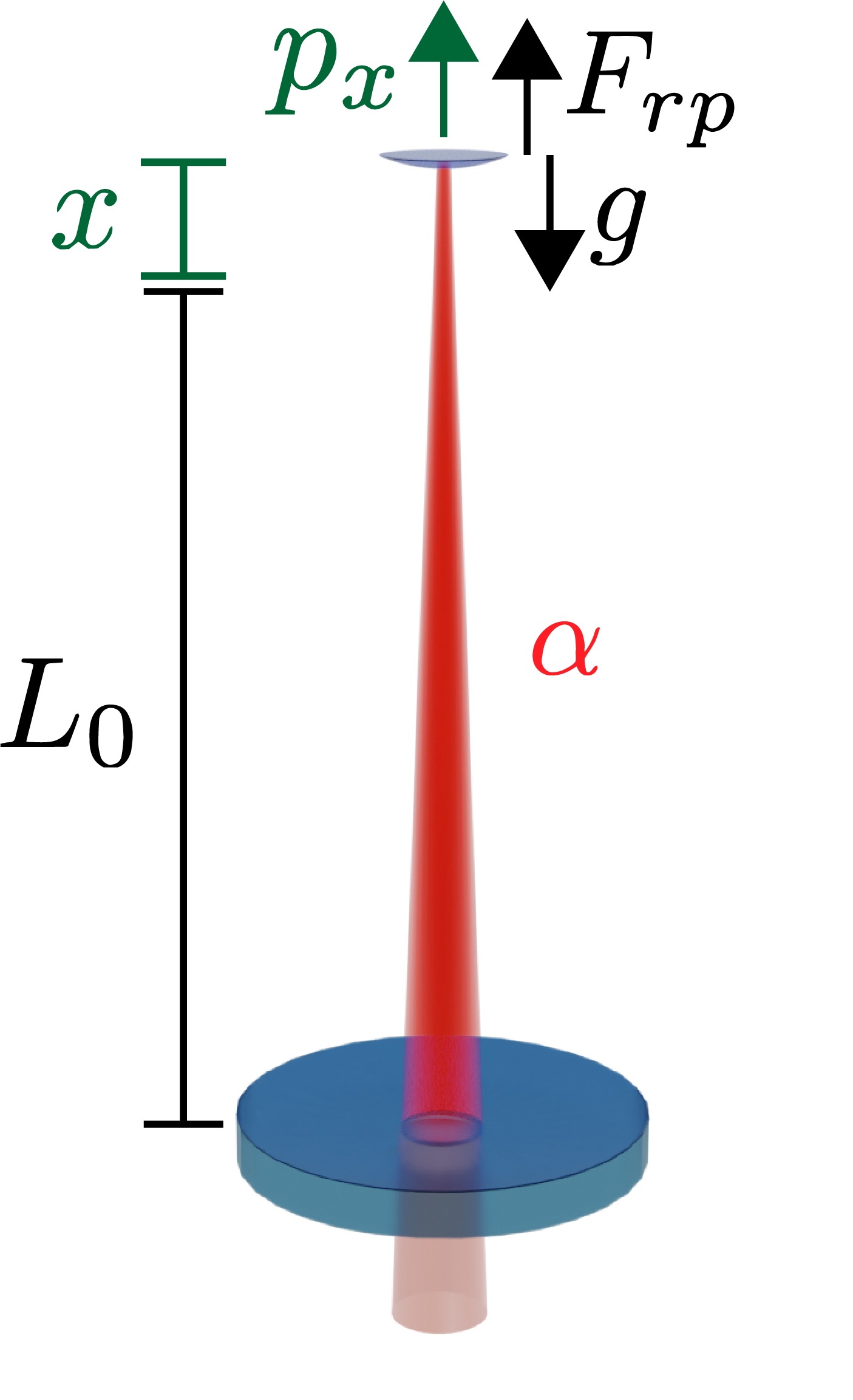}
  \caption{A vertically-oriented Fabry-P\'erot cavity with frequency $\omega_0$, natural length $L_0$, and intra-cavity field amplitude $\alpha$. The upper mirror is displaced a distance $x$ and has momentum $p_x$. The mirror is pushed downwards by gravity $g$, and upwards by radiation pressure force $F_{rp}=\hbar G|\alpha|^2$.}\label{figSystem}
\end{figure}

The system is shown in \cref{figSystem}, and consists of a vertically oriented Fabry-P\'erot cavity where the bottom mirror is fixed and the upper levitates due to the radiation pressure force of the intra-cavity field. The cavity has frequency $\omega_0$ and natural length $L_0$, and the deviation of the upper mirror's position from $L_0$ is denoted by $x$. The intra-cavity field amplitude is $\alpha$. In what follows, we will often refer to the levitating mirror as simply the `mirror'.

We will often want to compute the values of various quantities. A typical set of parameters for this system is a cavity finesse of 2500, a length of 10 cm, and a mirror mass of 1 mg. We will take the input laser to be 1050 nm with a power of 2 watts, and assume the input coupling is half of the linewidth ($\kappa_i=\delta\omega/2$ in \cref{eq:BaseEquations}).

\subsection{Dimensionless equations of motion}\label{sec:MainNondimensionalisation}%{[{

Suppose the cavity is driven by a coherent laser field with frequency $\omega_{\alpha}$. Moving to a reference frame rotating with the laser frequency, the system is then described by the equations of motion \cite[\S III.C]{Aspelmeyer2014} (where a dot denotes a derivative with respect to time $t$)
\begin{equation}\label{eq:BaseEquations}
  \begin{aligned}
    \dot{x}      &= \frac{p_x}{m}, \\
    \dot{p}_x    &= -mg+\hbar G|\alpha|^2, \\
    \dot{\alpha} &= i(\Delta_{\alpha}+G x)\alpha - \frac{\delta\omega}{2}\alpha+\sqrt{\frac{\kappa_i P_{\alpha}}{\hbar\omega_0}}.
  \end{aligned}
\end{equation}
Here $m$ is the mass of the mirror, with $x$ and $p_x$ its position and momentum respectively, and $g$ the gravitational acceleration. In the second equation the $\hbar G|\alpha|^2$ term represents the radiation pressure force on the mirror due to the cavity field, with $G$ the linearised optomechanical coupling \cite[\S III.B]{Aspelmeyer2014}. The input field of power $P_{\alpha}$ is detuned $\Delta_{\alpha}=\omega_0-\omega_{\alpha}$ relative to the cavity resonance, with input coupling $\kappa_i$. The cavity has linewidth $\delta\omega$, and $\hbar$ is the reduced Planck's constant.

\begin{table}
  \begin{tabular}{| c | c |}
  \hline
  \textbf{Variable} & \textbf{Meaning} \\
  \hline
                    & \emph{Mirror} \\
  $x$               & Mirror height \\
  $p_x$             & Mirror momentum \\
  $m$               & Mirror mass \\
  $g$               & Gravitational acceleration \\
  \hline
                    & \emph{Cavity} \\
  $L_0$             & Cavity natural length \\
  $G$               & Linear optomechanical coupling $\omega_0/L$ \\
  $\omega_0$        & Cavity frequency \\
  $\kappa_i$        & Cavity input coupling \\
  $\delta\omega$    & Cavity linewidth \\
  \hline
                    & \emph{First input laser} \\
  $\alpha$          & Amplitude of first cavity field \\
  $P_{\alpha}$      & Input power of first input laser \\
  $\omega_{\alpha}$ & Frequency of first input laser \\
  $\Delta_{\alpha}$ & First laser detuning $\omega_0-\omega_{\alpha}$ \\
%                    & \emph{Parameters $\beta$/$P_{\beta}$/\ldots are analogous.} \\
  \hline
                   & \emph{Second input laser} \\
  $\beta$          & Amplitude of second cavity field \\
  $P_{\beta}$      & Input power of second input laser $\beta$ \\
  $\omega_{\beta}$ & Frequency of second input laser $\beta$ \\
  $\Delta_{\beta}$ & Second laser detuning $\omega_0-\omega_{\beta}$ \\

  \hline
                   & \emph{Photothermal effects} \\
  $\zeta$          & Photothermal strength \\
  $\gamma$         & Photothermal relaxation rate \\
  \hline
  \end{tabular}
  \caption{Physical parameters used in the model.}\label{tab:Parameters}
\end{table}

To simplify the analysis, we nondimensionalise the equations of motion. What follows is a brief discussion of the results, for more details see \cref{sec:ANondimensionalisation}. We will introduce a natural length scale $\ell$, frequency $\nu$, and cavity amplitude $A$, defined as
\begin{equation}\label{eq:IdealNDScales}
  \ell=\frac{\delta\omega/2}{G},\;\nu=\sqrt{\frac{\hbar GA^2}{m\ell}},\;A=\frac{1}{\delta\omega/2}\sqrt{\frac{\kappa_iP_{\alpha}}{\hbar\omega_0}}.
\end{equation}
The parameter $\ell$ is the spatial displacement of the mirror for the cavity to have a detuning of one half-linewidth, and $A$ gives the maximum cavity amplitude: $0<|\alpha|^2<|A|^2$. The timescale of the upper mirror's oscillation frequencies is given by $\nu$, which is half the optical spring frequency at one lindwidth's displacement. Because of this we will often refer to $\nu$ as the `mechanical timescale' and $\delta\omega/2$ as the `optical timescale'.

In terms of these natural scales we define the dimensionless dynamical variables (denoted by a tilde):
\begin{equation}\label{eq:IdealNDSubstitutions}
  \tilde{\tau}=\nu t,\; \tilde{x}=\frac{x}{\ell},\; \tilde{p}=\frac{p}{m\ell\nu},\;\tilde{\alpha}=\frac{\alpha}{A}.
\end{equation}
Because of the chosen normalisation, we will have $0<|\tilde{\alpha}|^2<1$. Furthermore the mirror variables $\tilde{x}$ and $\tilde{p}_x$ will typically be of order $1$, as if $\tilde{x}\gg 1$ the cavity detuning will be far from resonance so the optical force will approach zero.

We will also require three dimensionless parameters:
\begin{equation}
  \tilde{g}=\frac{g}{\ell\nu^2},\;\tilde{\varepsilon}=\frac{\nu}{\delta\omega/2},\;\tilde{\Delta}_{\alpha}=\frac{\Delta_{\alpha}}{\delta\omega/2}.
\end{equation}
The simplest is $\tilde{\Delta}_{\alpha}$, which is merely the detuning rescaled in terms of the cavity linewidth. More interesting is the `effective gravity' $\tilde{g}$, which equals (simplifying $G=\omega_0/L_0$ \cite[\S III.B]{Aspelmeyer2014})
\begin{equation}
  \tilde{g}=\frac{mgL_0(\delta\omega/2)^2}{\kappa_iP_{\alpha}}.
\end{equation}
In the numerator are quantities which when increased make levitation more difficult, namely mass, gravity, cavity linewidth, and the natural length of the cavity. In the denominator we have the coupled input power.

Finally there is the parameter $\tilde{\varepsilon}$, which compares the response timescale of the cavity $\delta{\omega}/2$ to our natural timescale $\nu$: 
\begin{equation}
  \tilde{\varepsilon}=\frac{1}{L_0(\delta\omega/2)^{5/2}}\sqrt{\frac{\kappa_iP_{\alpha}\omega_0}{m}}
\end{equation}
To understand the dependence on cavity length it is helpful to expand $\delta\omega=(\pi c)/(L_0\mathcal{F})$, where $c$ is the speed of light and $\mathcal{F}$ the cavity finesse:
\begin{equation}
  \tilde{\varepsilon}=L_0^{3/2}\left(\frac{2\mathcal{F}}{\pi c}\right)^{5/2}\sqrt{\frac{\kappa_iP_{\alpha}\omega_0}{m}}
\end{equation}
Typically the intra-cavity field responds very quickly to changes in the position of the mirror, and so $\tilde{\varepsilon}$ will be very small. For the parameters mentioned in \cref{sec:MainAdiabatic} we have $\tilde{\varepsilon}\approx 1/6$, and if the cavity length is decreased to 1 cm this becomes $\tilde{\varepsilon}\approx 1/60$. We will exploit this separation of timescales to perform a perturbative expansion around $\tilde{\varepsilon}\approx 0$, which will simplify the dynamics to the point where an analytic approach is tractable. Numerical simulation will show that this picture is still reasonably accurate, even for values of $\tilde{\varepsilon}\approx 1/5$.

With these definitions, and letting primes denote derivatives with respect to $\tilde{\tau}$, the equations of motion are
\begin{equation}\label{eq:NDBaseEquations}
  \begin{aligned}
    \tilde{x}'      &= \tilde{p}_x, \\
    \tilde{p}_x'    &= -\tilde{g}+|\tilde{\alpha}|^2, \\
    \tilde{\alpha}' &= \left[i(\tilde{\Delta}_{\alpha}+\tilde{x})\tilde{\alpha}-\tilde{\alpha}+1\right]/\tilde{\varepsilon}.
  \end{aligned}
\end{equation}
From the differential equation for $\tilde{p}_x'$, we see that levitation can only occur in the regime $0<\tilde{g}<1$. Throughout the rest of this section we will understand this system by analysing it using successively weaker approximations.

\begin{table}
  \begin{tabular}{| c | c |}
    \hline
    \textbf{Variable} & \textbf{Meaning} \\
    \hline
                            & \emph{Natural scales} \\
    $\ell$                  & Natural length $\frac{\delta\omega/2}{G}$ \\
    $\nu$                   & Mechanical frequency $\sqrt{G^2P_{\alpha}\kappa_i/m\omega_0(\delta\omega/2)^3}$ \\

    \hline
                            & \emph{Cavity and Mirror} \\
    $\tilde{x}$             & $x/\ell$ \\
    $\tilde{p}_x$           & $p_x/m\ell\nu$ \\
    $\tilde{\sigma}$  & $\sqrt{\tilde{g}^{-1}-1}$ \\
    $\tilde{g}$             & Effective gravity $\left(mgL_0(\delta\omega/2)^2\right)/(\kappa_iP_{\alpha})$ \\
    $\tilde{\varepsilon}$   & Ratio of timescales $\nu/(\delta\omega/2)$ \\
    $\tilde{\chi}$          & $\tilde{x}+\tilde{\Delta}_{\alpha}$ \\

    \hline
                            & \emph{First input laser} \\
    $\tilde{\alpha}$        & $\alpha/A$ \\
    $A$                     & Cavity max amplitude $(\delta\omega/2)^{-1}\sqrt{(\kappa_iP_{\alpha})/(\hbar\omega_0)}$\\
    $\tilde{\Delta}_{\alpha}$ & $\Delta_{\alpha}/(\delta\omega/2)$ \\

    \hline
                            & \emph{Second input laser} \\
    $\tilde{\beta}$           & $\beta/(\tilde{B}A)$ \\
    $\tilde{B}$               & Ratio of max amplitudes of $\beta/\alpha$: $\sqrt{P_{\beta}/P_{\alpha}}$ \\
    $\tilde{\Delta}_{\beta}$  & $\Delta_{\beta}/(\delta\omega/2)$ \\
    $\tilde{\Delta}_{\beta\alpha}$ & $\tilde{\Delta}_{\beta}-\tilde{\Delta}_{\alpha}$ \\

    \hline
                            & \emph{Photothermal effects} \\
    $\tilde{\zeta}$           & $(\hbar\omega_0A^2\zeta)/(2\ell L_0/c)l$ \\
    $\tilde{\gamma}$          & $\gamma/\nu$ \\
    \hline
  \end{tabular}
  \caption{Natural parameters used in the model.}\label{tab:NonDimParameters}
\end{table}

%}]}
\subsection{Adiabatic limit}\label{sec:MainAdiabatic}%{[{
The optical timescale will typically be much faster than the mechanical one. We can exploit this to perform an `adiabatic approximation' and assume that for any given $\tilde{x}$, the light field will immediately reach its steady state given by $\tilde{\alpha}'=0$. Formally this corresponds to the limit $\tilde{\varepsilon}\rightarrow 0$, see \cref{sec:AExpandingEpsilon} for a detailed derivation. In this case we have
\begin{equation}\label{eq:AdiabaticAmplitude}
  |\tilde{\alpha}|^2 = \frac{1}{1+(\tilde{\Delta}_{\alpha}+\tilde{x})^2},
\end{equation}
giving us the reduced equations of motion:
\begin{equation}\label{eq:AdiabaticIdealEOM}
  \begin{aligned}
    \tilde{x}'    &= \tilde{p}_x, \\
    \tilde{p}_x'  &= -\tilde{g}+\frac{1}{1+(\tilde{\Delta}_{\alpha}+\tilde{x})^2}.
  \end{aligned}
\end{equation}
In \cref{sec:AAdiabatic} we show that these equations describe a classical particle moving in a potential
\begin{equation}\label{eq:DimPotential}
  \tilde{V}(\tilde{x})=\tilde{g}(\tilde{\Delta}_{\alpha}+\tilde{x})-\arctan\left(\tilde{\Delta}_{\alpha}+\tilde{x}\right),
\end{equation}
with total dimensionless energy
\begin{equation}\label{eq:AdiabaticEnergy}
  \tilde{\mathcal{E}}(\tilde{x},\tilde{p}_x)=\frac{\tilde{p}_x^2}{2}+\tilde{V}(\tilde{x}).
\end{equation}

\begin{figure}
  \subfloat[Slices of the dimensionless potential\label{fig:DimPotential}]{%
    \includegraphics[width=\figwidth\columnwidth]{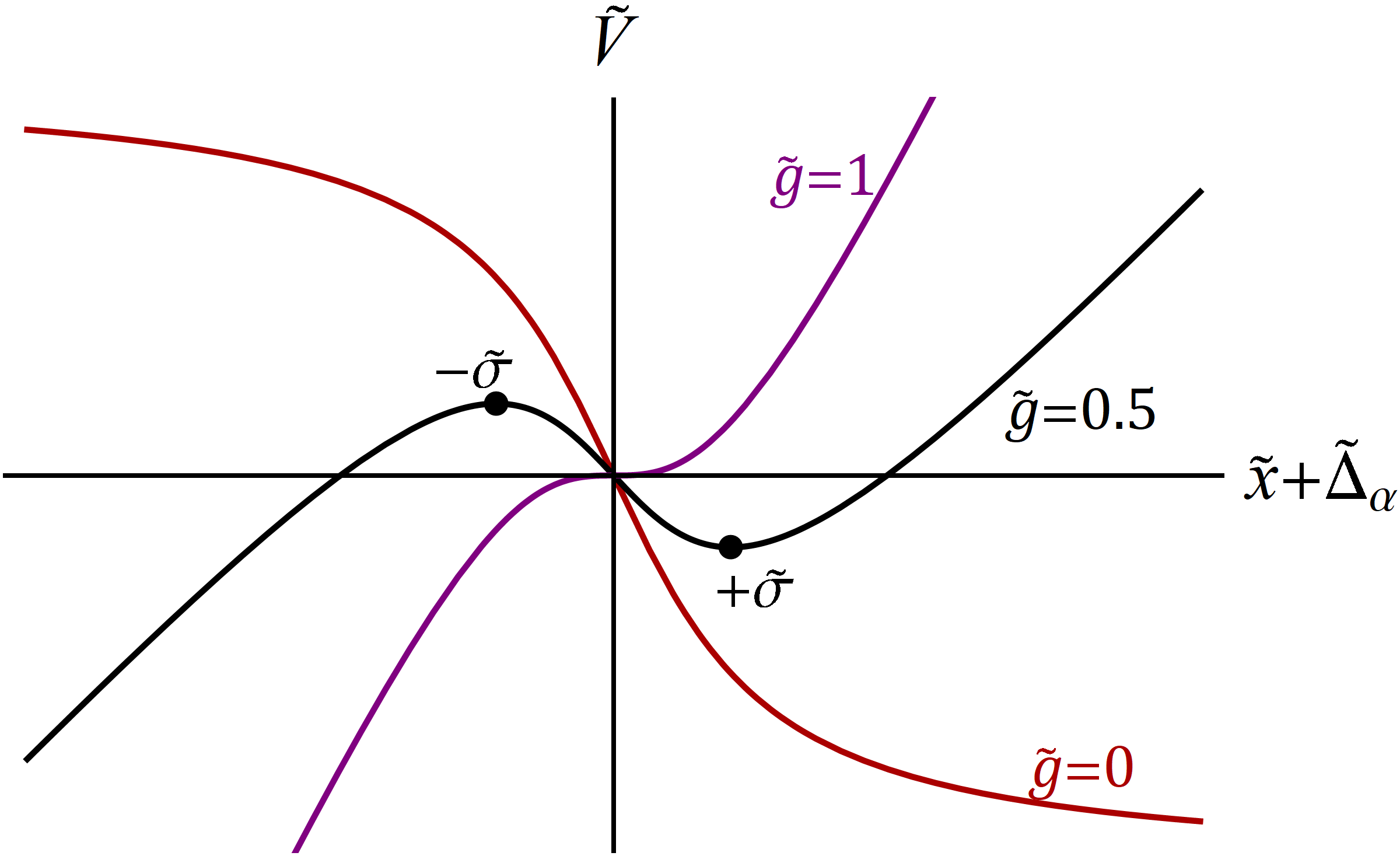}
  } \\
  \subfloat[Full dimensionless potential\label{fig:DimPotentialFull}]{%
    \includegraphics[width=\figwidth\columnwidth]{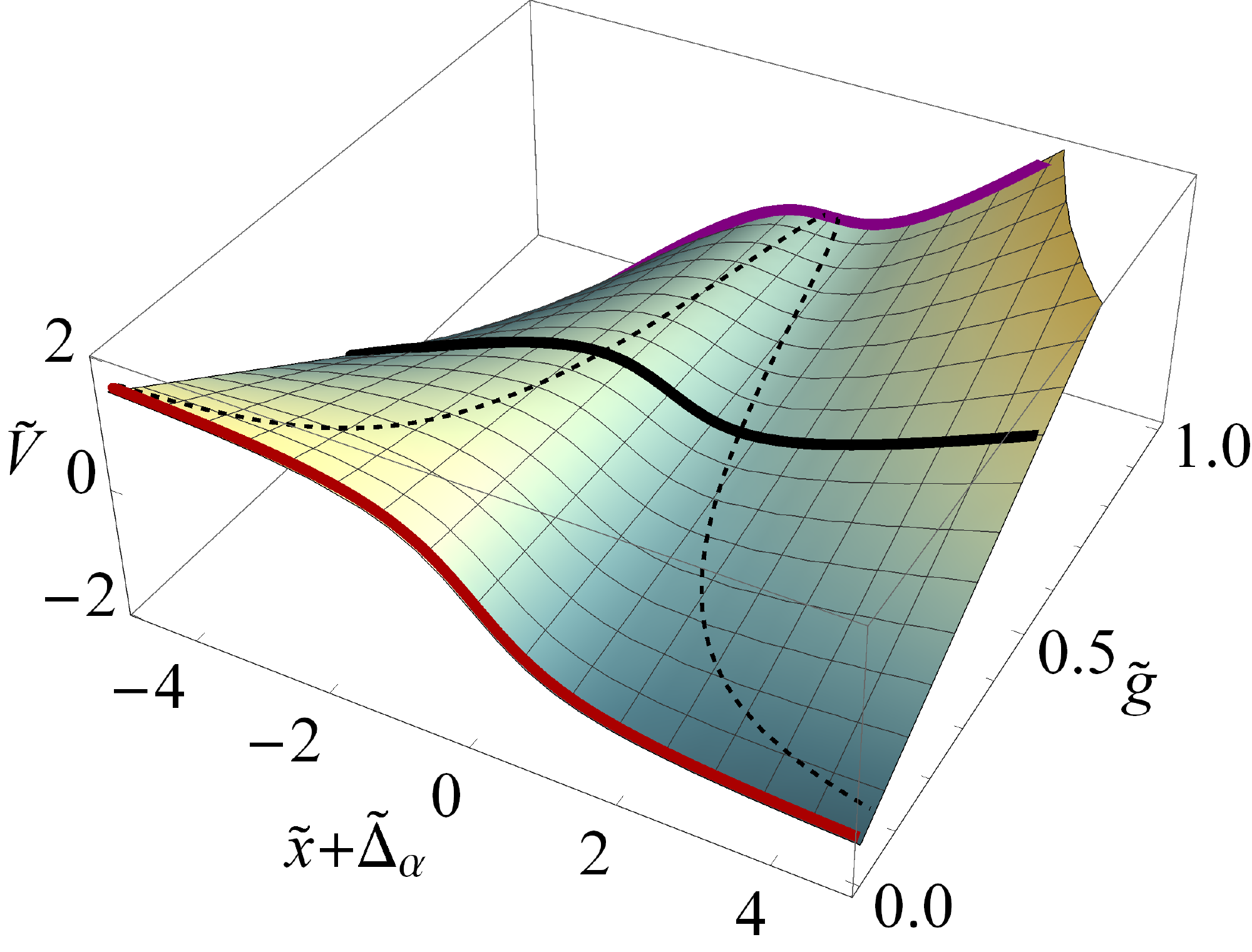}
  } \\
  \subfloat[Frequency of oscillation\label{fig:OscillationFrequency}]{%
    \includegraphics[width=\figwidth\columnwidth]{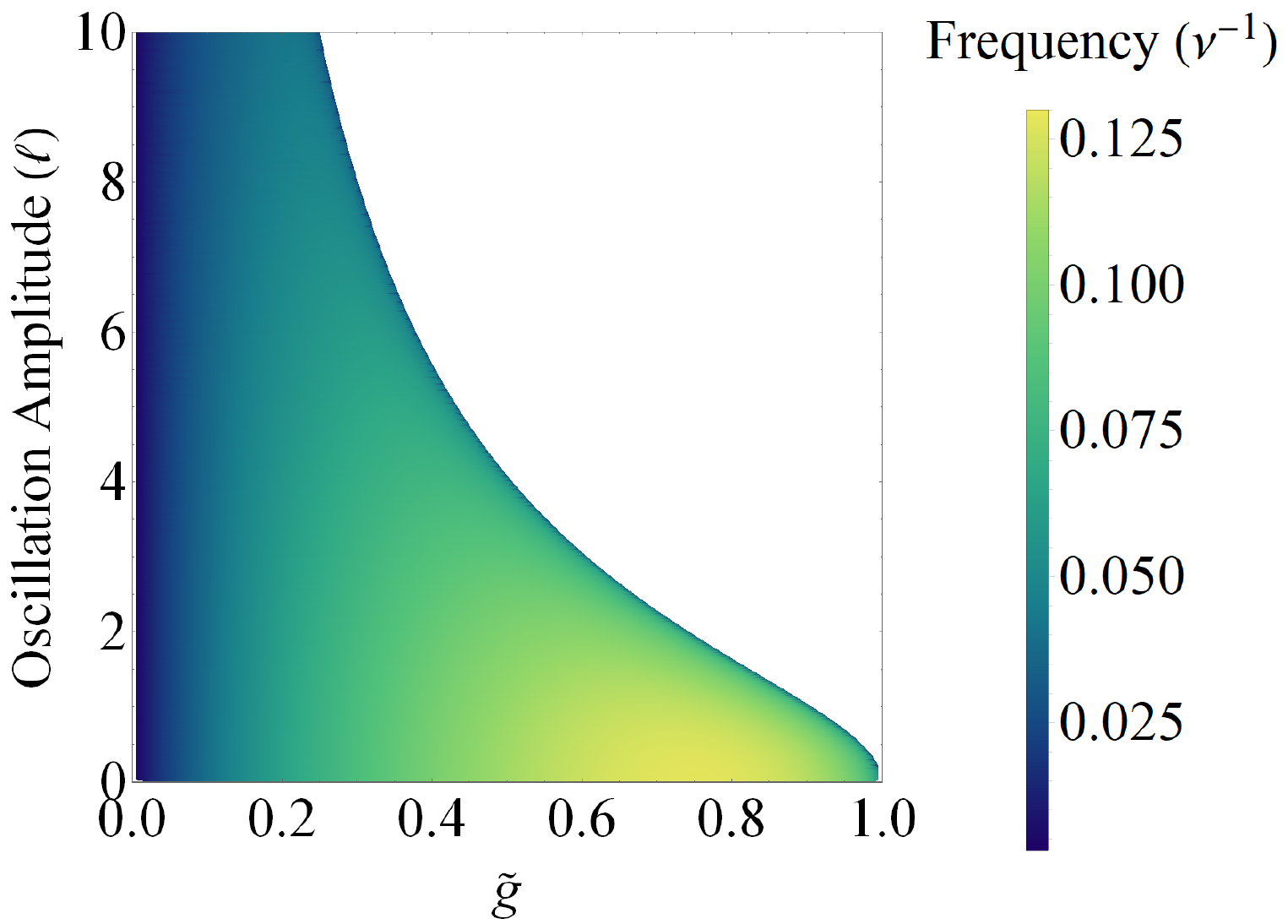}
  } \\
  \caption{\protect\subref{fig:DimPotential} The dimensionless potential \cref{eq:DimPotential}, plotted for various values of $\tilde{g}$. The two black dots represent the equilibrium points at $\pm\tilde{\sigma}$ for $\tilde{g}=0.5$. \protect\subref{fig:DimPotentialFull} The dimensionless potential over all values of $\tilde{g}$. The solid lines correspond to the values of $\tilde{g}$ plotted in \protect\subref{fig:DimPotential}, and the dashed lines show the equilibria at $\pm\tilde{\sigma}$. \protect\subref{fig:OscillationFrequency} The frequencies of oscillation divided by $\tilde{\nu}$, as a function of $\tilde{g}$ and amplitude scaled by $\tilde{\ell}$. The white region represents oscillation amplitudes which are larger than the width of the potential well for that value of $\tilde{g}$, and hence impossible.}
\end{figure}

We plot the potential $\tilde{V}(\tilde{x})$ in \cref{fig:DimPotential} for various values of $\tilde{g}$. This has two equilibrium points at 
\begin{equation}
  \tilde{x}_{\pm}^s= -\tilde{\Delta}_{\alpha}\pm\tilde{\sigma},
\end{equation}
where we have defined
\begin{equation}\label{eq:SteadyStateSigma}
  \tilde{\sigma}=\sqrt{\tilde{g}^{-1}-1}.
\end{equation}
The point $\tilde{x}^s_-$ is unstable, while $\tilde{x}^s_+$ allows for quasi-stable oscillations. Small $\tilde{g}$ gives a wide and shallow potential well, which morphs into a point of inflection as $\tilde{g}$ approaches its maximum value of $1$.

We detail in \cref{sec:AAdiabatic} how to use $\tilde{V}(\tilde{x})$ to calculate the possible frequencies of oscillation of the mirror, which will depend on the amplitude of oscillation since the well is anharmonic. The range of possible amplitudes decreases as $\tilde{g}$ increases, as this corresponds to the well shortening. We plot the frequencies in \cref{fig:OscillationFrequency}. Note that apart from amplitude, the frequency of oscillation depends solely on the dimensionless parameter $\tilde{g}$, and we observe that frequency decreases as the amplitude increases.

Finally, up to present we have been assuming that the detuning $\tilde{\Delta}_{\alpha}$ is static. In practice the system is often characterised by sweeping the detuning over a range of values: $\tilde{\Delta}_{\alpha}(\tilde{\tau})=\tilde{\Delta}_{\alpha0}+\tilde{s}\tilde{\tau}$, for $\tilde{\Delta}_{\alpha0}$, $\tilde{s}$ constants. Assuming the mirror begins resting on some solid support at $\tilde{x}=0$, we show in \cref{sec:AAdiabatic} that if the detuning is swept from negative to positive the mirror exhibits temporary oscillations, while if the detuning begins positive and decreases, the mirror will be lifted off the support if and only if the scan speed $\tilde{s}$ is less than the critical value
\begin{equation}
  \tilde{s}_c=2\sqrt{\arctan(\tilde{\sigma})-\tilde{\sigma}\tilde{g}}.
\end{equation}
%}]}
\subsection{Dynamics of the light field}\label{sec:MainDynamics}%{[{
We will now relax the adiabatic approximation, and consider the first-order perturbation to \cref{eq:AdiabaticIdealEOM} induced by the dynamics of the light field. In \cref{sec:AExpandingEpsilon} we derive the first-order correction 
\begin{equation}\label{eq:FirstOrderEOM}
  \begin{aligned}
    \tilde{x}'    &= \tilde{p}_x, \\
    \tilde{p}_x'  &= -\tilde{g}+\frac{1}{1+\left(\tilde{x}+\tilde{\Delta}_{\alpha}\right)^2}+\tilde{\varepsilon}\frac{4(\tilde{p}_x+\tilde{s})(\tilde{x}+\tilde{\Delta}_{\alpha})}{\left(1+(\tilde{x}+\tilde{\Delta}_{\alpha})^2\right)^3}.
  \end{aligned}
\end{equation}
Recall that in the adiabatic limit the mirror oscillated in the dimensionless potential without any loss of energy. If we compute the derivative of $\tilde{\mathcal{E}}$, hereafter referred to as the `heating rate', from \cref{eq:AdiabaticEnergy} with respect to the first-order equations \cref{eq:FirstOrderEOM}, we find (setting $\tilde{s}=0$ for simplicity)
\begin{equation}\label{eq:DampingRate}
  \frac{d\tilde{\mathcal{E}}}{d\tilde{\tau}}=\tilde{\varepsilon}\frac{4\tilde{p}_{\tilde{x}}^2(\tilde{x}+\tilde{\Delta}_{\alpha})}{\left(1+(\tilde{x}+\tilde{\Delta}_{\alpha})^2\right)^3}.
\end{equation}
The dynamics of the light field thus alters the energy of the system, and the sign of $\tilde{x}+\tilde{\Delta}_{\alpha}$ determines whether damping (negative) or anti-damping (positive) occurs. This corresponds to the usual optomechanical picture of sideband-heating (cooling) in the blue (red-detuned) input-laser regime \cite[\S VA]{Aspelmeyer2014}. Let us look again at the potential in \cref{fig:DimPotential}, and consider a small oscillation around the potential well minimum. Since the oscillations lie to the right of the axis, $\tilde{x}+\tilde{\Delta}_{\alpha}$ is positive; there will be anti-damping and the amplitude of the oscillation will grow. Eventually the oscillations will be so large that the mirror crosses the axis into the damping region, however as this is much narrower than the anti-damping region, we can expect that over one oscillation the anti-damping will dominate, and the mirror will eventually cross over $-\tilde{\sigma}$ and leave the trap. The effect of the dynamics of the light field is thus to render unstable all oscillations within the trapping well.

\begin{figure}
  \subfloat[Comparing dynamics with approximation for $\tilde{\varepsilon}=1/100$\label{fig:EvaluatingApproxTrace}]{%
    \includegraphics[width=\columnwidth]{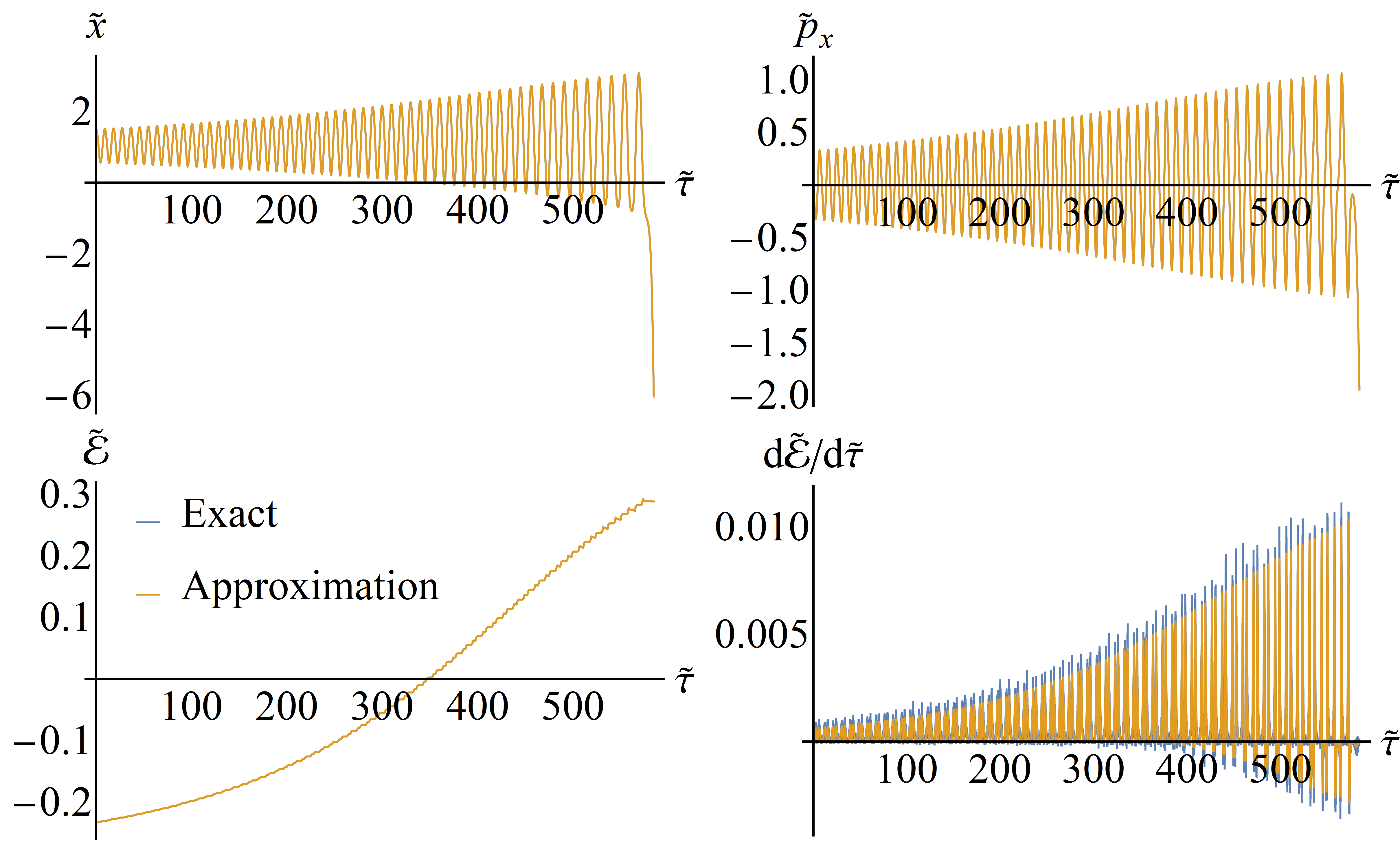}
  } \\
  \subfloat[Zoomed-in comparison of heating rates for $\varepsilon=1/100$\label{fig:EvaluatingApproxZoomed}]{%
    \includegraphics[width=0.8\columnwidth]{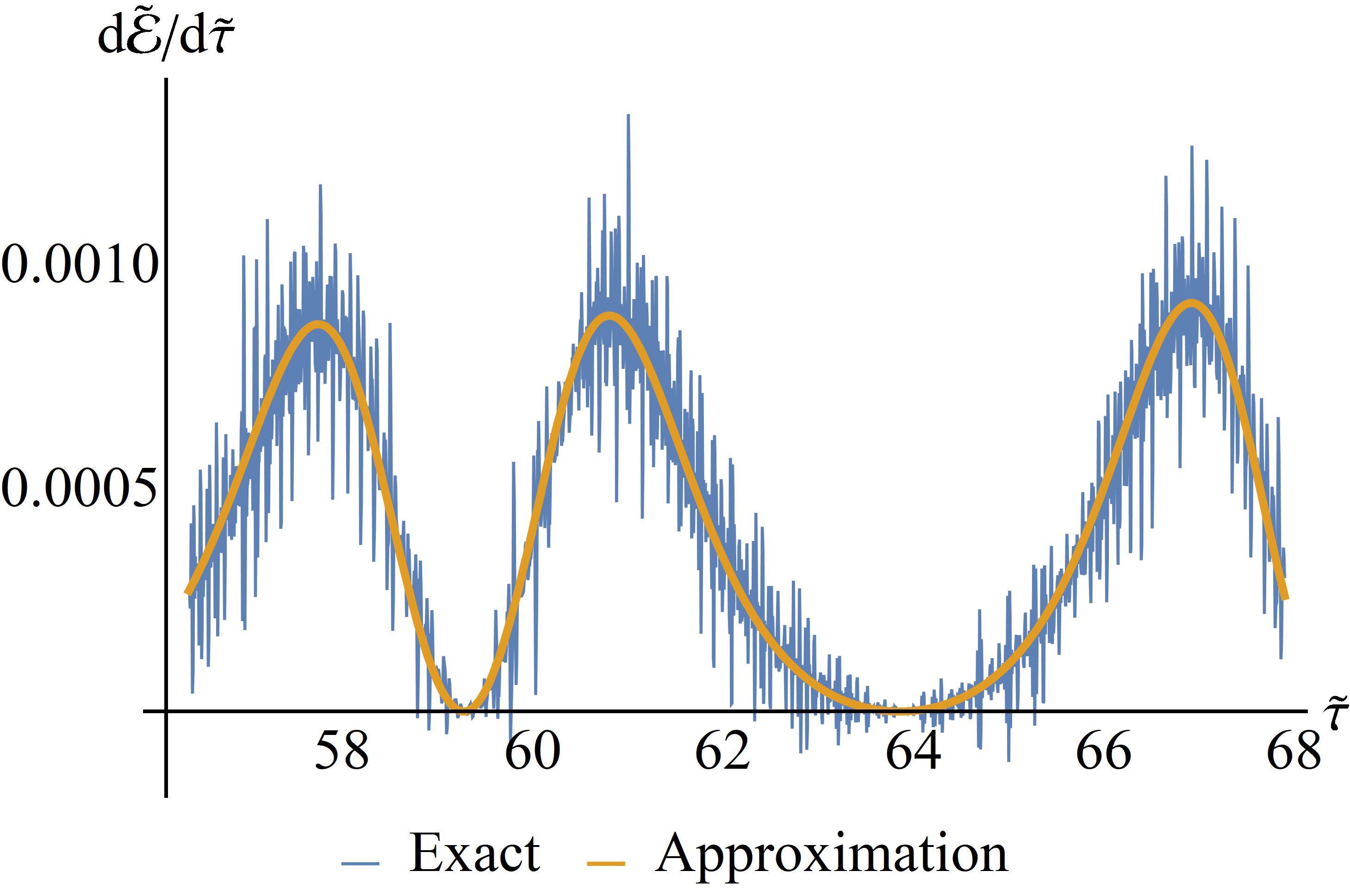}
  } \\
  \subfloat[Comparing dynamics with approximation for $\tilde{\varepsilon}=1/10$\label{fig:EvaluatingApproxSmallEta}]{%
    \includegraphics[width=\columnwidth]{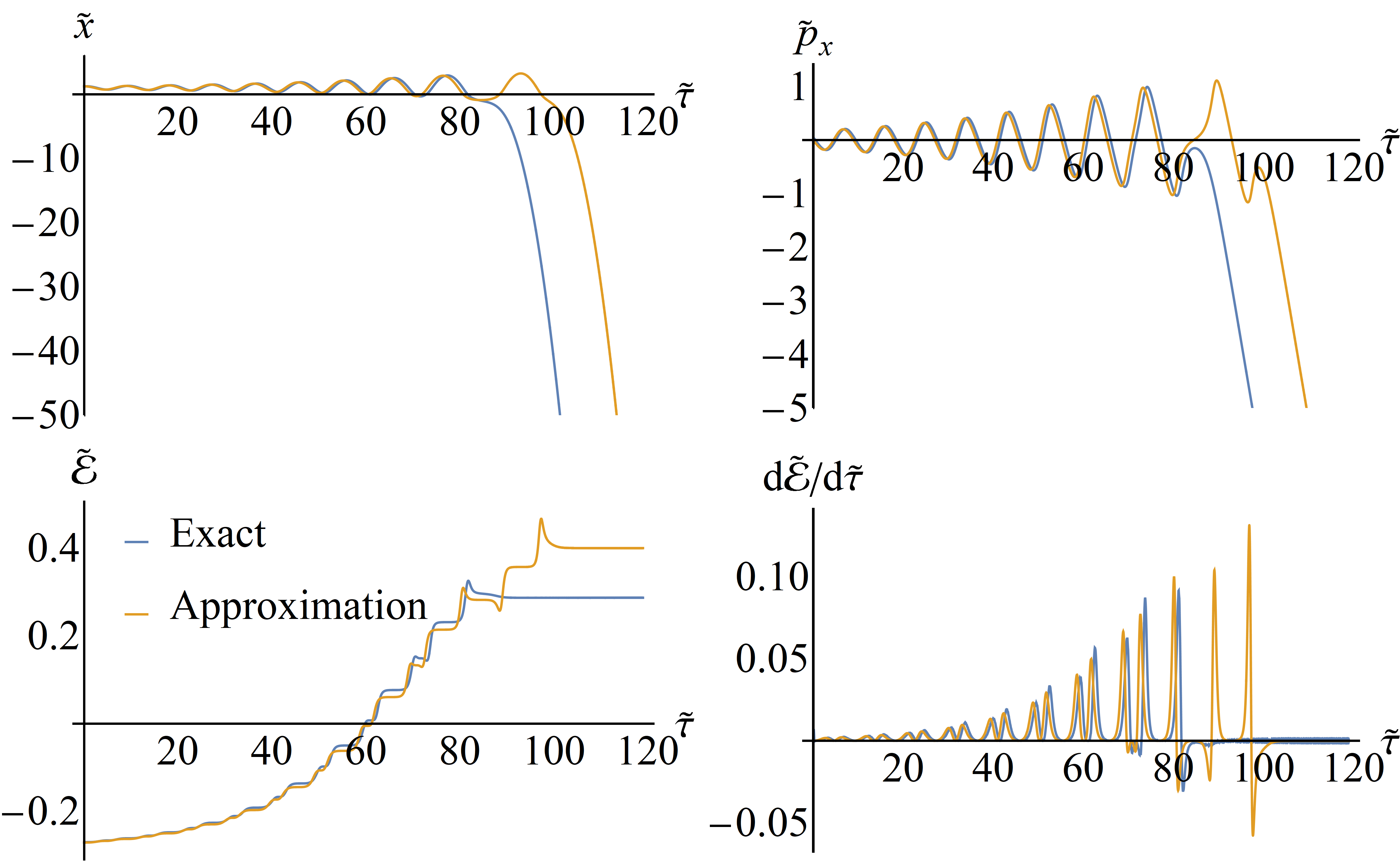}
  } \\
  \caption{\protect\subref{fig:EvaluatingApproxTrace} Comparison of the first-order approximation \cref{eq:FirstOrderEOM} (orange) with the full system \cref{eq:NDBaseEquations} (blue), with parameters $(\tilde{g},\tilde{\varepsilon},\tilde{\Delta}_{\alpha})=(0.5,1/100,0)$. The mirror starts off slightly perturbed from equilibrium, and exhibits growing oscillations until it falls out of the trapping well. The traces for $\tilde{x}$, $\tilde{p}_x$, and the energy $\tilde{\mathcal{E}}$ perfectly coincide, while we can see slight discrepancies in the heating rate $\mathrm{d}\tilde{\mathcal{E}}/{\mathrm{d}}\tilde{\tau}$. \protect\subref{fig:EvaluatingApproxZoomed} Zooming in to the heating rate $d\tilde{\mathcal{E}}/d\tilde{t}$ we see that the actual equations of motion predict very fast behaviour, the average of which is well described by the approximation. \protect\subref{fig:EvaluatingApproxSmallEta} For $\tilde{\varepsilon}=1/10$ the actual dynamics eventually diverge from those of the approximation, however they are still qualitatively similar.}\label{fig:EvaluatingApproximation}
\end{figure}

Finally, we investigate in \cref{fig:EvaluatingApproximation} how accurate the approximation made in this section is. We see in \cref{fig:EvaluatingApproxTrace} that for a value of $\tilde{\varepsilon}=1/100$ there is no visible difference between the approximation and full simulation for $\tilde{x}$ and $\tilde{p}_x$. Zooming in on the heating rate in \cref{fig:EvaluatingApproxZoomed}, this is found to be highly oscillatory due to the rapid dynamics of the cavity field, but the average value is well-described by the approximation. For $\tilde{\varepsilon}=1/10$ shown in \cref{fig:EvaluatingApproxSmallEta} we get a divergence, however the dynamics are still qualitatively similar. The approximation may thus still be effective for analysing qualitative features of dynamics, even in reasonably large $\tilde{\varepsilon}$ regimes.

%}]}
\subsection{Visualising the motion}\label{sec:MainVisualisation} %{[{
\begin{figure}
  \subfloat[Motion in the effective potential for $\tilde{\varepsilon}=1/5$\label{fig:VisualisationMotionInPotential}]{%
    \includegraphics[width=0.9\columnwidth]{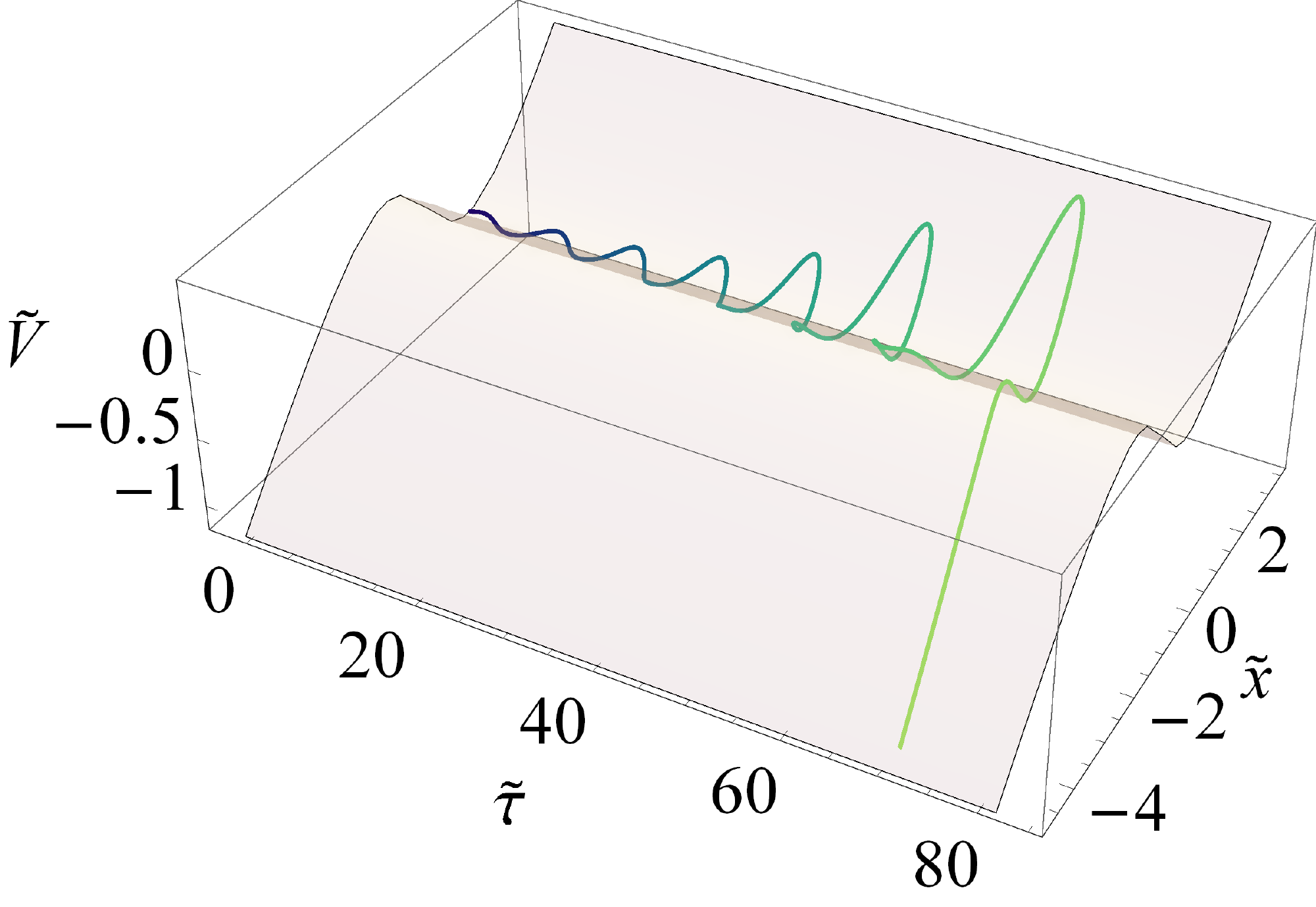}
  } \\
  \subfloat[Motion in phase space for $\tilde{\varepsilon}=1/5$\label{fig:VisualisationPhaseSpace}]{%
    \includegraphics[width=0.9\columnwidth]{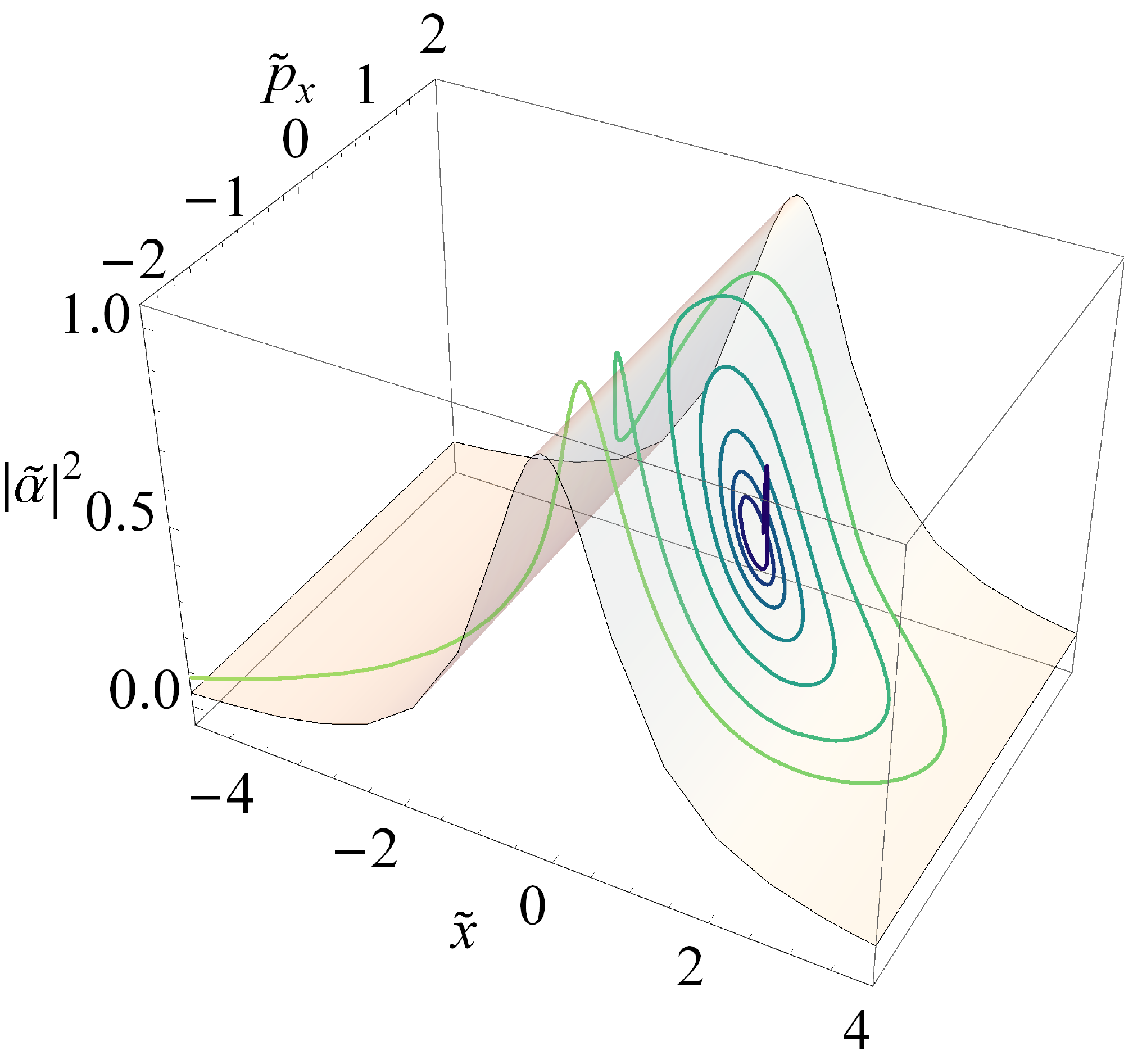}
  } \\
  \caption{\protect\subref{fig:VisualisationMotionInPotential} A plot of $\tilde{x}$ vs $\tilde{t}$ for $(\tilde{g},\tilde{\Delta}_{\alpha},\tilde{\varepsilon})=(0.5,0,1/5)$, where the trajectory is coloured according to time $\tilde{\tau}$ (beginning at blue and ending at yellow), and shown moving in the effective potential $\tilde{V}(\tilde{x})$ defined in \cref{eq:DimPotential}. The mirror exhibits growing oscillations in the trapping region, until eventually it falls out. \protect\subref{fig:VisualisationPhaseSpace} The same trajectory now plotted in phase space. The dynamics rapidly collapse onto the Lorentzian-shaped adiabatic manifold given by \cref{eq:AdiabaticAmplitude}, then exhibit growing oscillations until they fall out of the trap.}\label{fig:Visualisation}
\end{figure}

The simpler picture of the adiabatic and first-order approximations provides us with an intuitive visualisation of the system dynamics. Consider the trace for position $\tilde{x}$ in \cref{fig:EvaluatingApproxTrace}; the mirror exhibits growing oscillations until at some point it falls out of the trap. We can better understand this by plotting the mirror's motion in the potential $\tilde{V}(\tilde{x})$ in \cref{fig:VisualisationMotionInPotential} (simulated for a different value of $\tilde{\varepsilon}$ to \cref{fig:EvaluatingApproximation}). Now we can clearly see how the mirror is oscillating in a trapping region, and that the fall occurs once the oscillations take it out of the potential well. Note that we are simulating the full dynamics \cref{eq:NDBaseEquations} without any approximation.

In \cref{fig:VisualisationPhaseSpace} we plot the same trajectory in phase space. Regardless of starting point, the dynamics rapidly collapse onto the Lorentzian-shaped adiabatic manifold given by \cref{eq:AdiabaticAmplitude}, indicated by the orange surface. The system exhibits growing oscillations on the manifold, until eventually falling out of the trap. Thus while we have three dynamical variables, the separation of timescales means that the dynamics are effectively two-dimensional. 

The potential and adiabatic manifold shown in \cref{fig:Visualisation} are derived in the limit $\tilde{\varepsilon}\rightarrow 0$. Note however that the trajectory we simulate is for the somewhat large value of $\tilde{\varepsilon}=1/5$. Even at such values, our analysis can still provide us with an effective intuition for the system. The trajectory does briefly leave the manifold when passing over the maximum of the Lorentzian, but as $\tilde{\varepsilon}$ decreases these deviations will become much less noticeable. 

%}]}
%}]}
\section{Applications of Ideal Analysis}\label{sec:Application}%{[{
In this section we will apply the analysis of \cref{sec:MainAnalysis} to two scenarios of experimental relevance. Firstly we will look at passive sideband cooling, a commonly used technique in optomechanics, and show that it is much less effective for a levitating mirror system. Secondly we investigate photothermal effects of the mirror substrate. This is likely to become significant at the powers required for levitation, and we will show that this can have a stabilising or de-stabilising influence depending on the sign of the effect.

\subsection{Passive Cooling}\label{sec:ApplicationTwoLaser}%{[{
It is common in optomechanics to use a (possibly second) laser to passively damp the mirror's motion, a technique known as sideband cooling \cite[\S V.B]{Aspelmeyer2014}. Depending on the detuning of an input laser, it can either lead to a trapping force and anti-damping, opposing motion of the mirror but transferring energy to it from the optical field, or anti-trapping but damping, removing energy from the mirror while pushing it in the direction of motion. In the case of a levitated mirror the idea would be to use a strong primary laser to provide the trapping, and a weaker, detuned, ancillary laser to provide the damping. By choosing appropriate amplitudes and detunings for the two lasers, one aims to find a regime where the net trapping and damping effects dominate. In contrast to a normal optomechanical system however, we do not have a mechanical spring, and so all of the trapping force must come from the light field. This makes a difference, and while there do exist regions with simultaneous trapping and damping, the potential around these tends to be both shallow and flat.

Suppose we have a second laser incident on the cavity, leading to another mode with amplitude $\beta$. The equations of motion \cref{eq:BaseEquations} then become
\begin{equation}\label{eq:TwoLaserBaseEquations}
  \begin{aligned}
    \dot{x}      &= \frac{p_x}{m}, \\
    \dot{p}_x      &= -mg+\hbar G|\alpha|^2+\hbar G|\beta|^2, \\
    \dot{\alpha} &= i(\Delta_{\alpha}+G x)\alpha - \frac{\delta\omega}{2}\alpha+\sqrt{\frac{\kappa_i P_{\alpha}}{\hbar\omega_0}}, \\
    \dot{\beta} &= i(\Delta_{\beta}+G x)\beta - \frac{\delta\omega}{2}\beta+\sqrt{\frac{\kappa_i P_{\beta}}{\hbar\omega_0}},
  \end{aligned}
\end{equation}
where now $P_{\beta}$ and $\Delta_{\beta}$ are the power and detuing of the laser driving $\beta$. As before many of these parameters may be eliminated via nondimensionalisation, where now we will exploit the symmetry between $\alpha$, $\beta$. Without loss of generality we may take $P_{\alpha}\ge P_{\beta}$, and in analogy to \cref{eq:IdealNDSubstitutions} write
\begin{equation}
  \tilde{\beta}=\frac{\beta}{\tilde{B}A},
\end{equation}
where $\tilde{B}=\sqrt{P_{\beta}/P_{\alpha}}$ is between zero and one, and we will also have $|\tilde{\beta}|^2<1$. Defining $\tilde{\Delta}_{\beta\alpha}=\tilde{\Delta}_{\beta}-\tilde{\Delta}_{\alpha}$ and $\tilde{\chi}=\tilde{x}+\tilde{\Delta}_{\alpha}$, the non-dimensionalised equations are
\begin{equation}\label{eq:TwoLasersNDEOM}
  \begin{aligned}
    \tilde{\chi}'      &= \tilde{p}_x, \\
    \tilde{p}_x'    &= -\tilde{g}+|\tilde{\alpha}|^2 + \tilde{B}^2|\tilde{\beta}|^2, \\
    \tilde{\alpha}' &= \left[i\tilde{\chi}\tilde{\alpha}-\tilde{\alpha}+1\right]/\tilde{\varepsilon}, \\
    \tilde{\beta}' &= \left[i(\tilde{\chi}+\tilde{\Delta}_{\beta\alpha})\tilde{\beta}-\tilde{\beta}+1\right]/\tilde{\varepsilon}.
  \end{aligned}
\end{equation}
Inspecting the equation for $\tilde{p}_x'$, for there to be levitation we must have $\tilde{g}<1+\tilde{B}^2$.

We first analyse this in the adiabatic approximation as in \cref{sec:MainAdiabatic}, leading to an effective potential which is the sum of the potentials of the individual lasers:
\begin{equation}\label{eq:TwoLaserPotential}
  \tilde{V}(\tilde{\chi})= \tilde{g}\tilde{\chi}-\arctan\left(\tilde{\chi}\right)-\tilde{B}^2\arctan\left(\tilde{\chi}+\tilde{\Delta}_{\beta\alpha}\right).
\end{equation}
A first-order perturbation as in \cref{sec:MainDynamics} then gives us a heating rate
\begin{equation}\label{eq:TwoLaserHeating}
  \frac{d\tilde{\mathcal{E}}}{d\tilde{\tau}}=4\tilde{p}_x^2\tilde{\varepsilon}\left(\frac{\tilde{\chi}}{\left(1+\tilde{\chi}^2\right)^3}+\tilde{B}^2\frac{(\tilde{\chi}+\tilde{\Delta}_{\beta\alpha})}{\left(1+(\tilde{\chi}+\tilde{\Delta}_{\beta\alpha})^2\right)^3}\right).
\end{equation}

These equations can be used to find regions which are both trapping (minima of $\tilde{V}(\tilde{\chi})$) and damping (negative $d\tilde{\mathcal{E}}/d\tilde{\tau}$). Equivalently we could linearise the equations of motion and search for parameter regimes where the Jacobian has negative eigenvalues. Our approach however also allows us to visualise the width and shape of the trapping potential and (anti)damping. There are three dimensionless parameters to search over: $0\le\tilde{B}\le 1$, $0\le\tilde{g}\le1+\tilde{B}^2$, and $\tilde{\Delta}_{\beta\alpha}$, as $\tilde{\varepsilon}$ changes only the magnitude of the (anti)damping but not its sign. While there is no bound on the detuning, we restrict ourselves to $-10\le\tilde{\Delta}_{\beta\alpha}\le 10$.  The code and simulation results are available for download at ref. \cite{supplementaryGithub}.

\begin{figure}
  \subfloat[Simultaneous trapping and damping, $(\tilde{g},\tilde{B},\tilde{\Delta}_{\beta\alpha},\tilde{\varepsilon})\approx(0.37,0.47,-2.63,1/5)$\label{fig:TrappingCoolingWell}]{%
    \includegraphics[width=0.7\columnwidth]{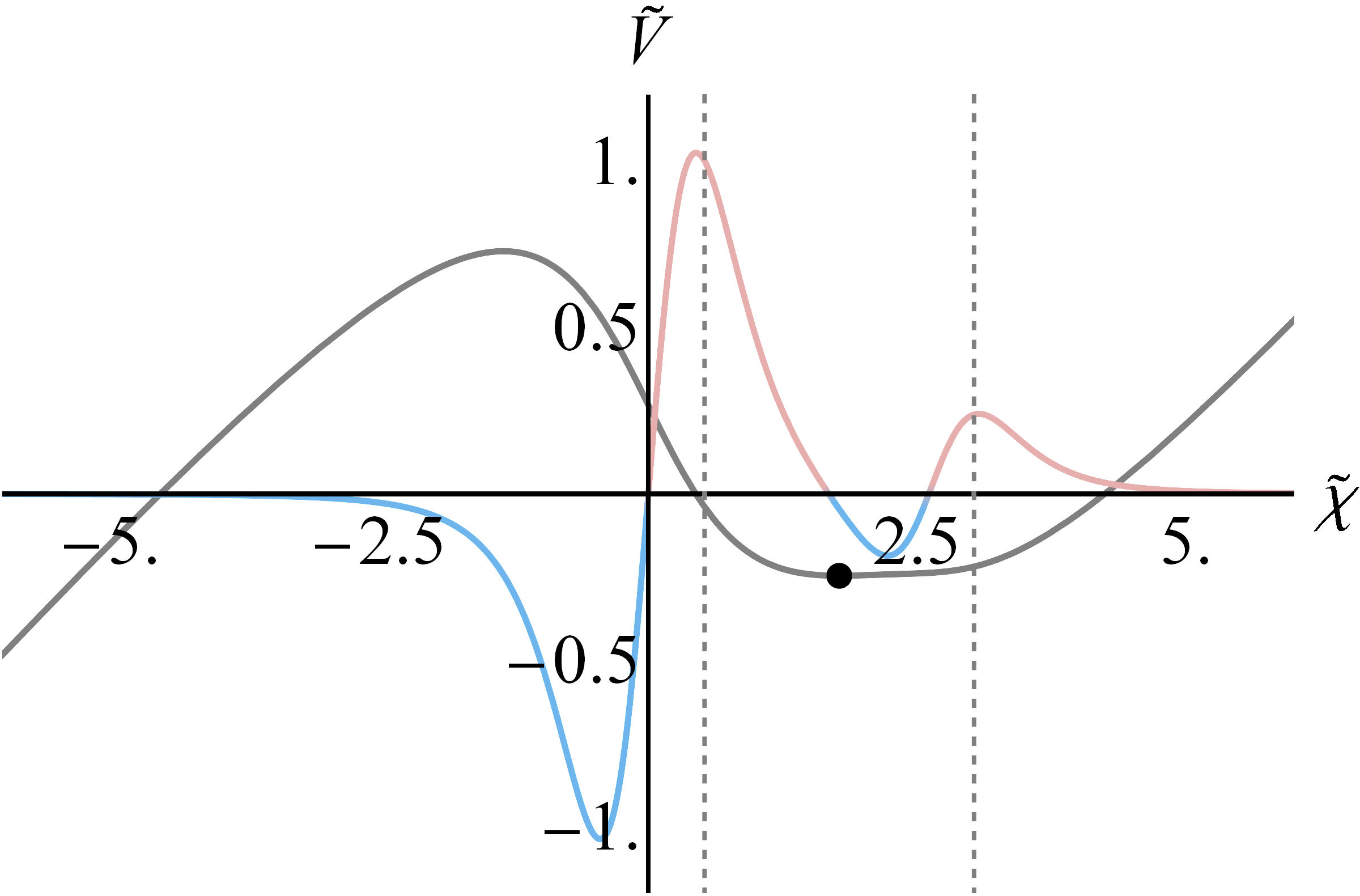}
  } \\
  \subfloat[Parameters at which there exists simultaneous trapping and damping, coloured by trapping width\label{fig:TrappingCoolingWidths}]{%
    \includegraphics[width=0.9\columnwidth]{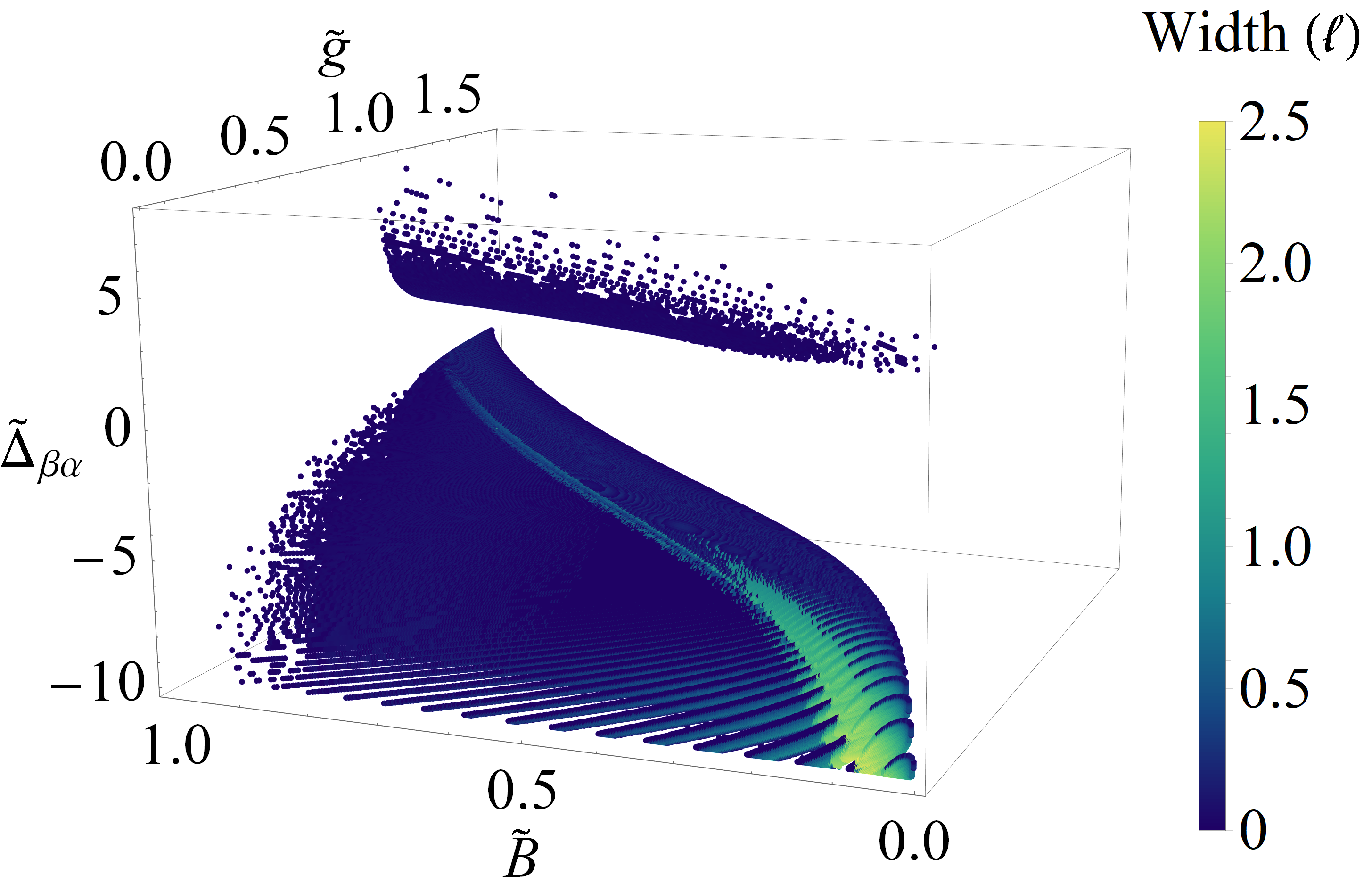}
  } \\
  \subfloat[Parameters at which there exists simultaneous trapping and damping, coloured by trapping area\label{fig:TrappingCoolingAreas}]{%
    \includegraphics[width=0.9\columnwidth]{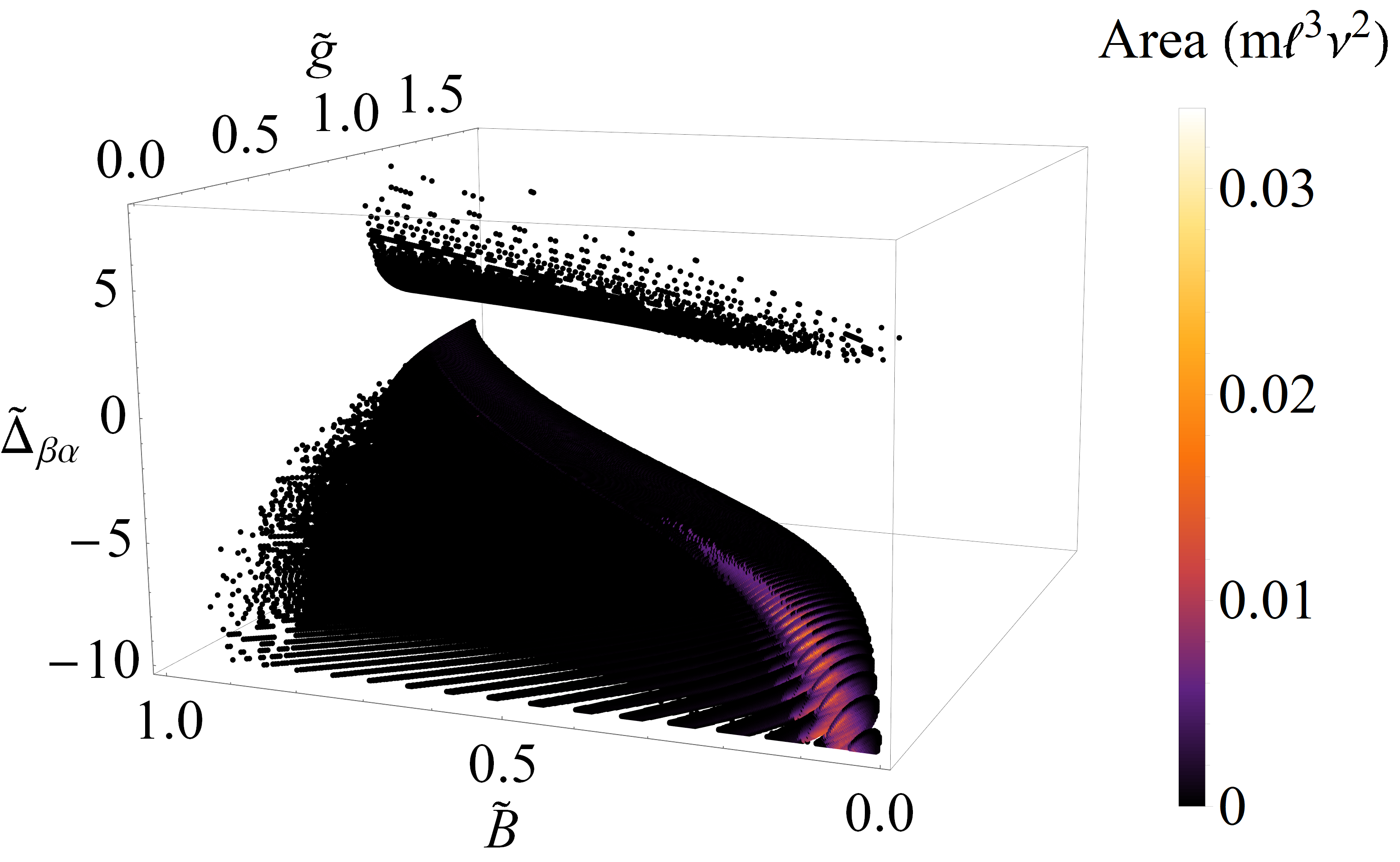}
  } \\
  \caption{\protect\subref{fig:TrappingCoolingWell} Regime with the greatest trapping and damping area. The black curve shows the effective two-laser potential \cref{eq:TwoLaserPotential} with the black dot the minimum, and the coloured line gives the effective heating rate \cref{eq:TwoLaserHeating} divided by $4\tilde{p}_x^2\tilde{\varepsilon}$, with blue for damping and red anti-damping. The two dashed vertical lines denote the two ends of the trapping region; trajectories originating in the trapping region with zero initial momentum will exhibit decaying oscillations towards the minimum. The distance between them is the `width' of this trap, and the vertical distance from the minimum to the intersection of the right dashed line with the potential is the `depth'. Multiplying the width by the depth then gives us the `area'. \protect\subref{fig:TrappingCoolingWidths} The parameters which allow for simultaneous trapping and damping regions map out a two-dimensional folded manifold in parameter space, with $\tilde{\varepsilon}=1/5$. Little difference is visible when $\tilde{\varepsilon}=1/100$ (although the (anti)damping rates will be much less). The points are coloured by the width of the region in units of $\ell$. \protect\subref{fig:TrappingCoolingAreas} The points are now coloured by the area of the trapping and damping region, found by multiplying the width by the depth.}\label{fig:TrappingCooling}
\end{figure}

One trapping and damping solution is plotted in \cref{fig:TrappingCoolingWell}. The dashed lines denotes the `trapping region', and trajectories beginning with these will exhibit decaying oscillations towards the minimum. As we can see, a trajectory beginning in a location where the optical field is anti-damping may still show net damping over one oscillation and be trapped, if for example it spends more time in the damping region over each period. The width of this trap is the space between the dashed lines ($\sim 1.3\;\mathrm{\ell}$), and the depth is the vertical distance between the minimum and the intersection of the rightmost dashed line with the potential ($\sim 0.03\;\mathrm{m \ell^2\nu^2}$), since if the mirror moves above this then it will leave the trapping region. The energy value the depth represents is very small, for the parameters in \cref{sec:MainAnalysis} this corresponds to the kinetic energy of a mirror with speed $10^{-5}\;\mathrm{ms^{-1}}$. However, note that from \cref{eq:DampingRate} the damping rate is proportional to momentum, and simulations show that an initial velocity at least ten times larger is still trapped and damped, though increasing by another factor of ten leads to the mirror escaping. 

In \cref{fig:TrappingCoolingWidths} we take $\tilde{\varepsilon}=1/5$ and plot the dimensionless parameters at which there exist regions of simultaneous trapping and damping. We searched over 500 values of $\tilde{g}$, 500 values of $\tilde{B}$, and $1000$ values of $\tilde{\Delta}_{\beta\alpha}$ within the aforementioned ranges, in a computation which took two days on a Google Cloud Compute Engine with 64 virtual CPUs. These form a folded, two-dimensional surface, where the banding comes from the numerical grid used. The colour of the points denotes the width of the trapping region. This was found by simulating the unapproximated equations of motion \cref{eq:TwoLasersNDEOM}, with $\tilde{\chi}$ beginning at increasingly large distances $\tilde{\chi}_0$ from the potential well minimum, and recording the largest distance such that both a positive and negative perturbation of that magnitude to the position would remain trapped. The initial conditions for the other dynamical variables were their steady states for the given initial value of $\tilde{\chi}_0$: $\tilde{p}_x=0$, $\tilde{\alpha}=1/(1-i\tilde{\chi}_0)$, and $\tilde{\beta}=1/\left(1-i(\tilde{\chi}_0+\tilde{\Delta}_{\beta\alpha})\right)$.  We can see that in general there are reasonably wide trapping and damping regions in the regime where both $\tilde{g}$ and $\tilde{B}$ approach zero, and $\tilde{\Delta}_{\beta\alpha}$ is strongly negatively detuned.

In \cref{fig:TrappingCoolingAreas} we colour the points by the `area', na\"ively found by multiplying the width by the depth of the well.  All of the trapping regions are very shallow, and so while robust against large changes in $\tilde{\chi}$, even a small change in momentum may be enough for the mirror to escape. We note that \cref{fig:TrappingCoolingWell} plotted the solution with the greatest trapping area, and other solutions display similar characteristics, with a broad flat region about the minimum and a somewhat shallow nature. Thus the experimental utility of sideband cooling for levitated systems is far from guaranteed, and a detailed investigation needs to be done to study the robustness of the parameter regimes identified against fluctuations of the cavity field or the mirror's thermal Brownian motion.

Finally, we repeated the simulations for $\tilde{\varepsilon}=1/100$ but there was little difference in the cooling widths and areas. The change in energy of the mirror's oscillations however, either damping or anti-damping, was much slower. This is to be expected, as (anti)damping arises from the dynamics of the light field, and as $\tilde{\varepsilon}\rightarrow 0$ these dynamics vanish.

%}]}
\subsection{Photothermal Effects}\label{sec:ApplicationPhotothermal}%{[{
So far we have assumed the mirror to be perfectly reflecting, which will not be true in practice. The intra-cavity field in this system must be very strong in order to provide levitation, and so even for a highly reflective mirror a substantial amount of optical energy is likely to be absorbed by the mirror substrate. This induces photothermal effects, where the mirror both deforms and undergoes a change in refractive index, resulting in a measurable change to the effective optical path length of the cavity. We can model this as \cite{Marino2006,Marino2007,Marino2011,Marino2013}
\begin{equation}
  \begin{aligned}
    \dot{x}       &= \frac{p_x}{m}, \\
    \dot{p}_x     &= -mg+\hbar G|\alpha|^2, \\
    \dot{z}       &= -\gamma\left(z+\zeta\frac{\hbar\omega_0|\alpha|^2}{2L_0/c}\right), \\
    \dot{\alpha}  &= i\left(\Delta_{\alpha}+G(x+z)\right)-\frac{\delta\omega}{2}\alpha+\sqrt{\frac{\kappa_iP_{\alpha}}{\hbar\omega_0}}.
  \end{aligned}
\end{equation}
The variable $z$ represents the change in optical path length due to photon absorption by the mirror, and may be positive or negative depending on the mirror substrate. The intra-cavity power is given by $\frac{\hbar\omega_0|\alpha|^2}{2L_0/c}$ (energy divided by photon travel time), and so $\zeta$ parameterises the rate at which intracavity power translates to a change in path length, with units length divided by power. The timescale of the effect is quantified by the photothermal relaxation rate $\gamma$.

As in \cref{sec:MainNondimensionalisation} we may derive non-dimensionalised equations of motion:
\begin{equation}\label{eq:FullNDEOM}
  \begin{aligned}
    \tilde{x}'        &= \tilde{p}_x, \\
    \tilde{p}_x'      &= -\tilde{g}+|\tilde{\alpha}|^2, \\
    \tilde{z}'        &= -\tilde{\gamma}\left[\tilde{z}+\tilde{\zeta}|\tilde{\alpha}|^2\right], \\
    \tilde{\alpha}'   &= \left[i(\tilde{\Delta}_{\alpha}+\tilde{x}+\tilde{z})\tilde{\alpha}-\tilde{\alpha}+1\right]/\tilde{\varepsilon},
  \end{aligned}
\end{equation}
where we have introduced dimensionless parameters
\begin{equation}
  \tilde{z}=\frac{z}{\ell},\;\tilde{\gamma}=\frac{\gamma}{\nu},\;\tilde{\zeta}=\frac{\hbar\omega_0 A^2\zeta}{(2L_0/c)\ell}.
\end{equation}
We will consider the regime $\tilde{\gamma}\ll 1$, which is true for many systems due to the slow rate of thermal expansion and contraction.  For example, a typical value of $\gamma$ between 10-100 Hz \cite{DeRosa2002} will make $\tilde{\gamma}\sim 10^{-5}-10^{-4}$ (for the parameters from \cref{sec:MainAnalysis}). The expansion coefficient $\zeta$ depends on the mirror substrate and the width and shape of the optical beam, a typical value is on the order of 10 picometers per watt \cite[\S III.C]{Konthasinghe2017} which corresponds to $\tilde{\zeta}\sim 30$.

We now analyse how the photothermal expansion affects the equilibrium states of the system. From $\tilde{x}'$ we clearly must have $\tilde{p}_x^s=0$.  The equation for $\tilde{p}_x'$ gives that $|\tilde{\alpha}^s|^2=\tilde{g}$, and so from $\tilde{z}'$ we find:
\begin{equation}
  \tilde{z}^s = -\tilde{\zeta}\tilde{g}.
\end{equation}
Thus the only contribution of the photothermal effect to the steady state is to add a constant shift to the detuning. As before the mirror position $\tilde{x}$ will have two equilibrium states 
\begin{equation}\label{eq:PTSteadyState}
  \tilde{x}^s_{\pm}=-\tilde{\Delta}_{\alpha}^s\pm\tilde{\sigma},
\end{equation}
with $\tilde{\sigma}$ from \cref{eq:SteadyStateSigma} and effective detuning
\begin{equation}
  \tilde{\Delta}_{\alpha}^s=\tilde{\Delta}_{\alpha}-\tilde{\zeta}\tilde{g}.
\end{equation}

\begin{figure}
  \includegraphics[width=0.48\columnwidth]{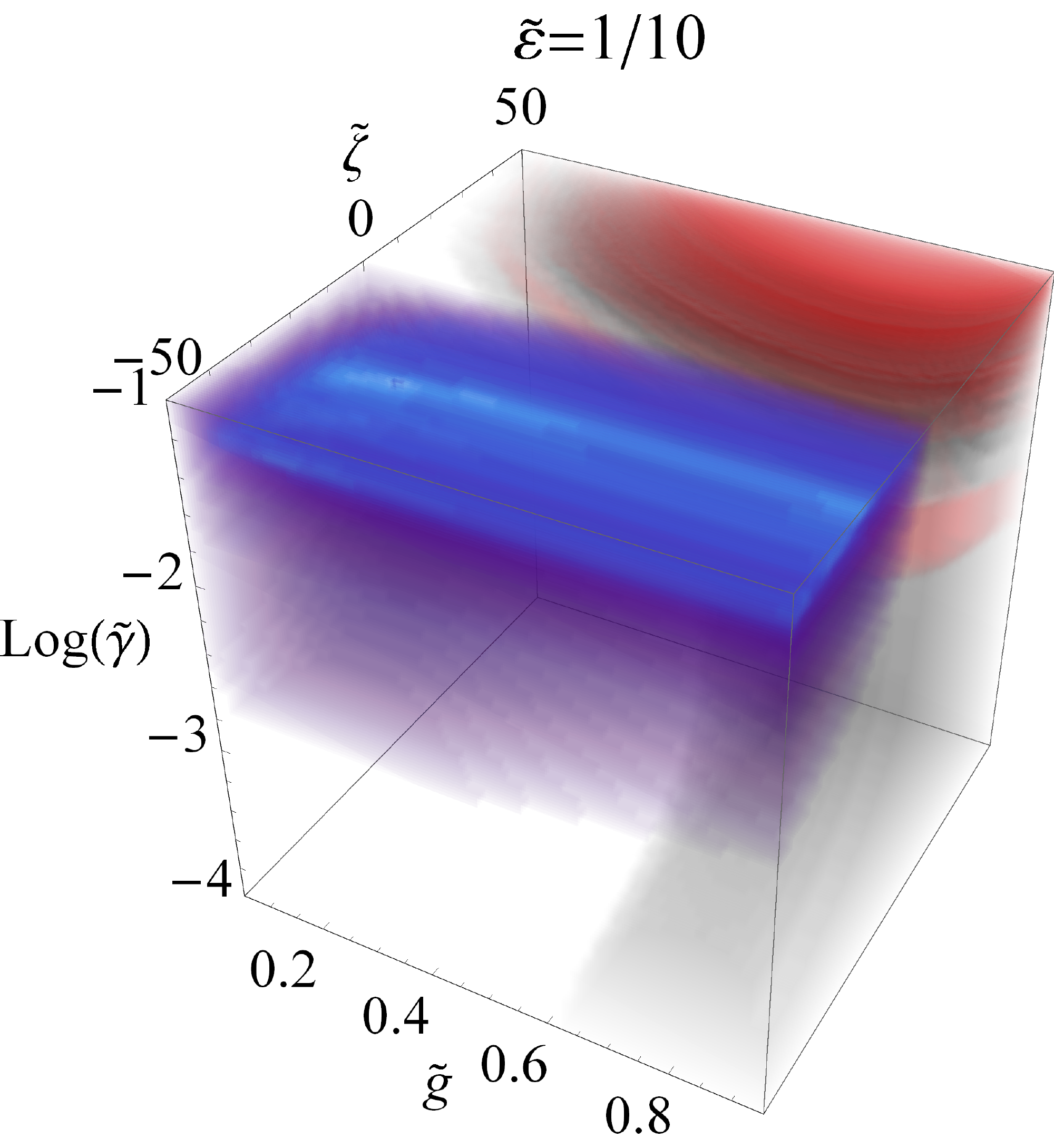}
  \includegraphics[width=0.48\columnwidth]{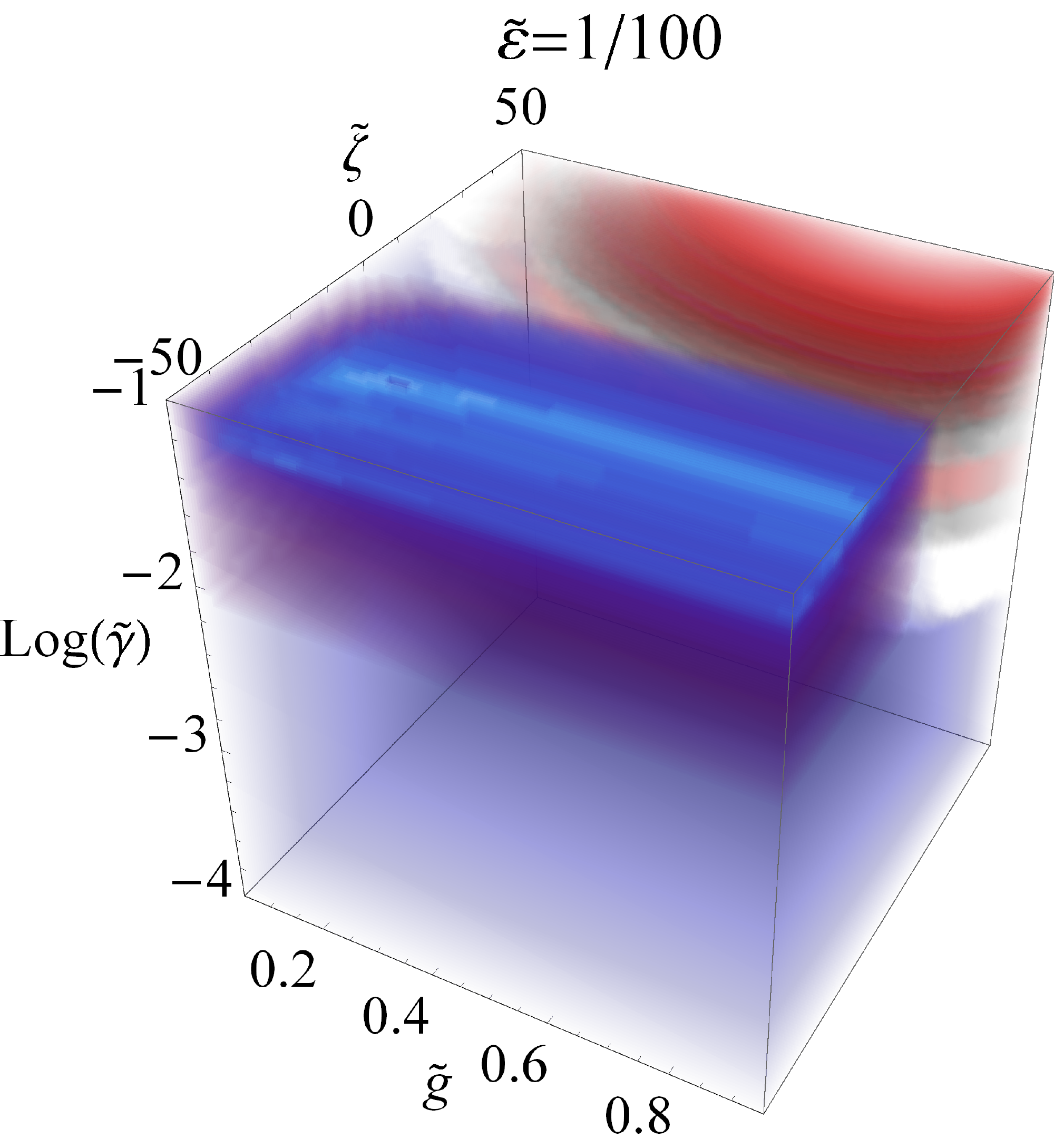}
  \includegraphics[width=0.48\columnwidth]{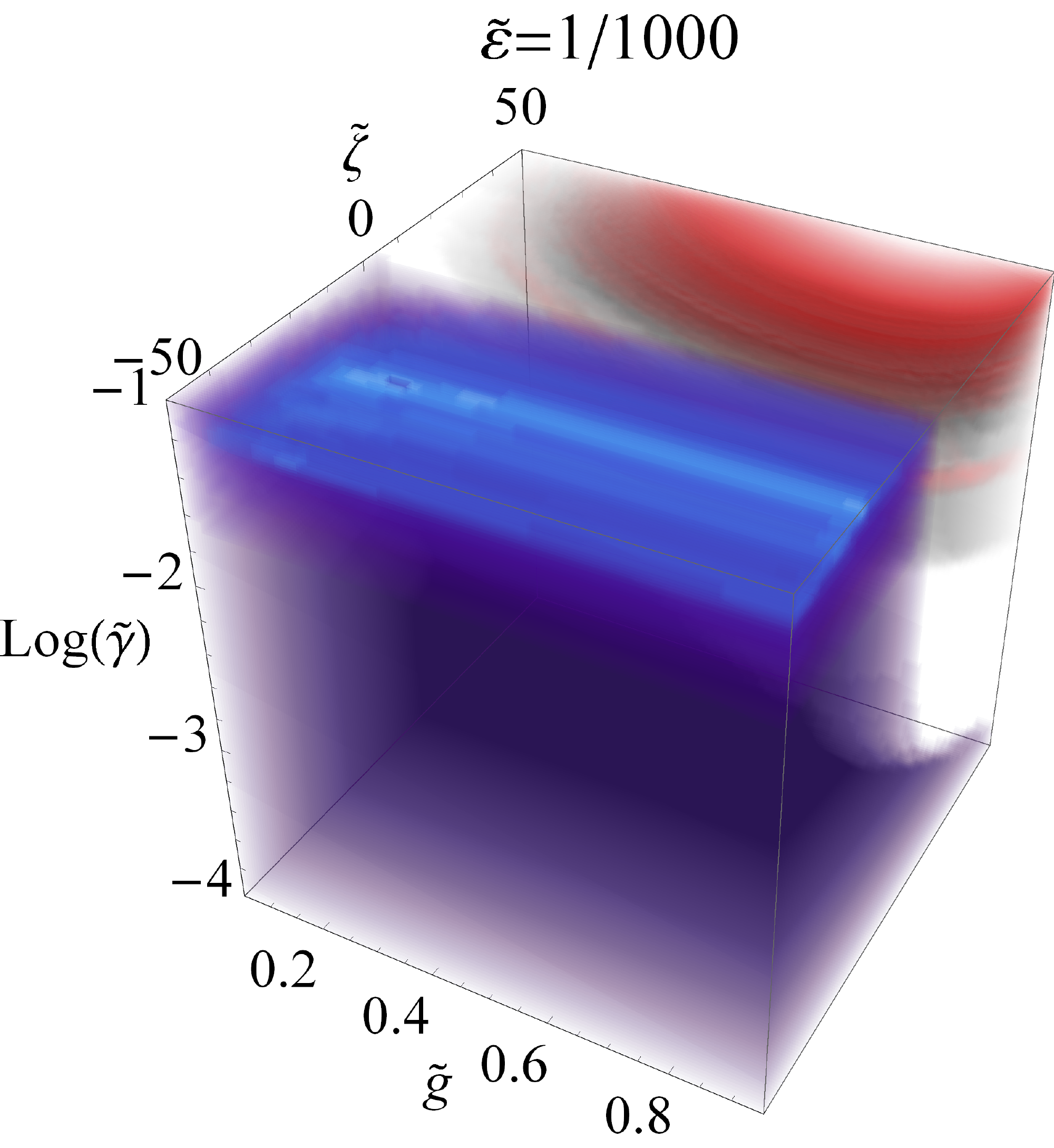}
  \includegraphics[height=4.5cm]{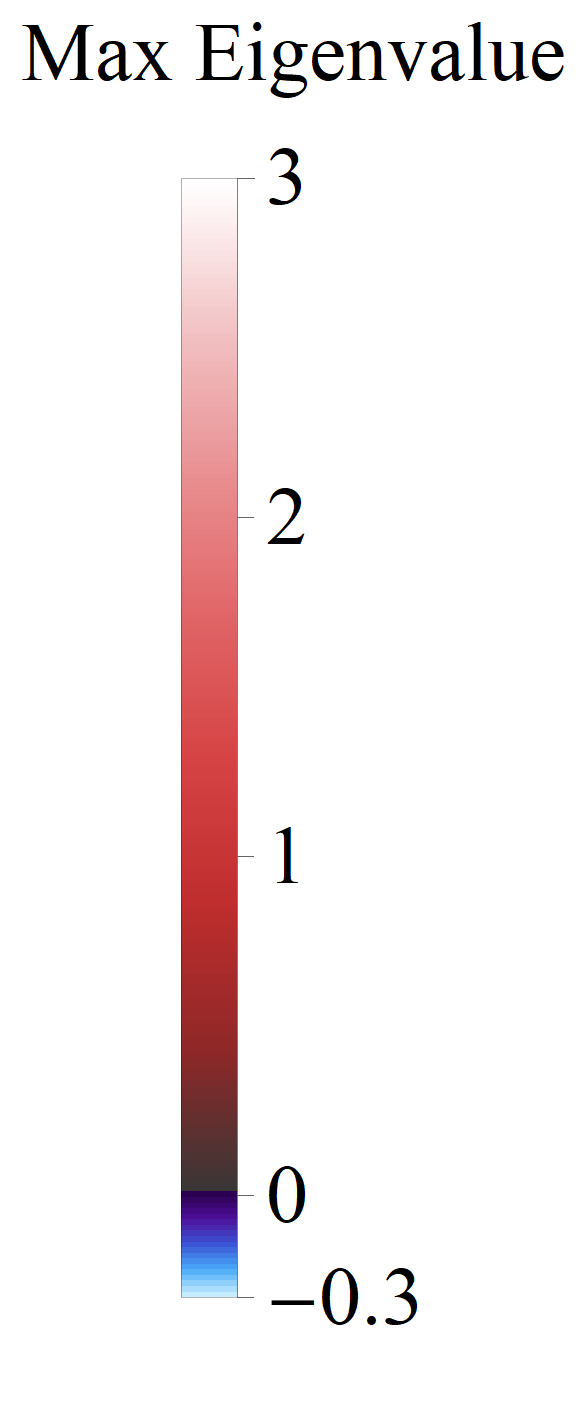}
  \caption{We take $\tilde{\gamma}=10^{-4}$ and compute the maximum eigenvalue of the linearised system at the potential-well minimum $\tilde{x}^s_+$ for various values of $\tilde{\varepsilon}$. Positive and negative values are coloured red and blue respectively, and the opacity is proportional to the magnitude of the eigenvalue (for negative eigenvalues a value of $-0.3$ is fully opaque, while for the positive eigenvalues a value of $3$ is opaque). For large $\tilde{\varepsilon}$ there is competition between photothermal damping and radiation pressure anti-damping, while as $\tilde{\varepsilon}$ decreases and optical effects become less significant the sign of the eigenvalues relies on $\tilde{\zeta}$ alone.}\label{fig:PhotothermalEigenvalues}
\end{figure}

\begin{figure}
  \subfloat[State space of system with photothermal expansion\label{fig:PhotothermalStateSpace}]{%
    \includegraphics[width=0.8\columnwidth]{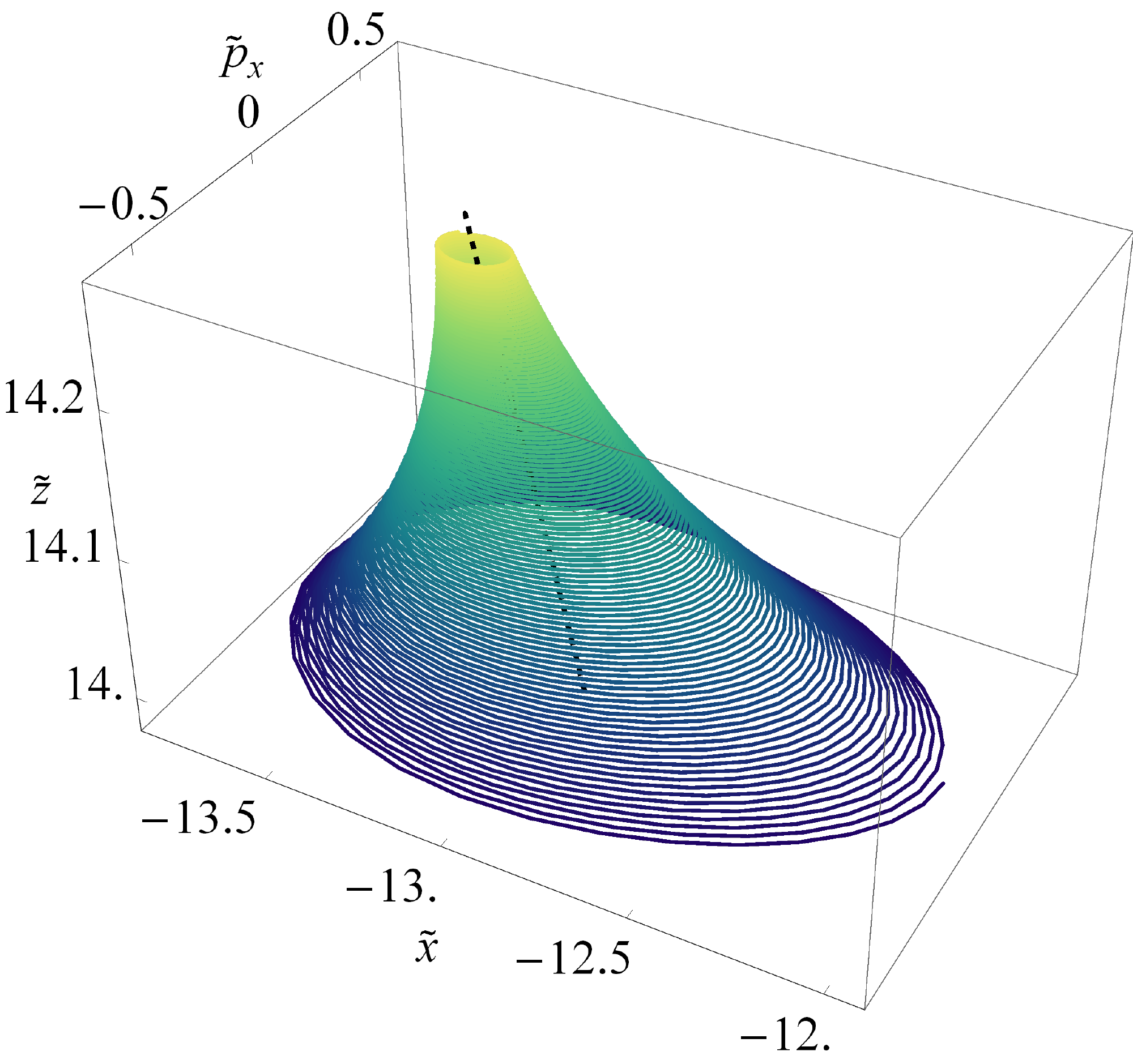}
  } \\
  \subfloat[Motion in potential with photothermal damping\label{fig:PhotothermalPotential}]{%
    \includegraphics[width=0.9\columnwidth]{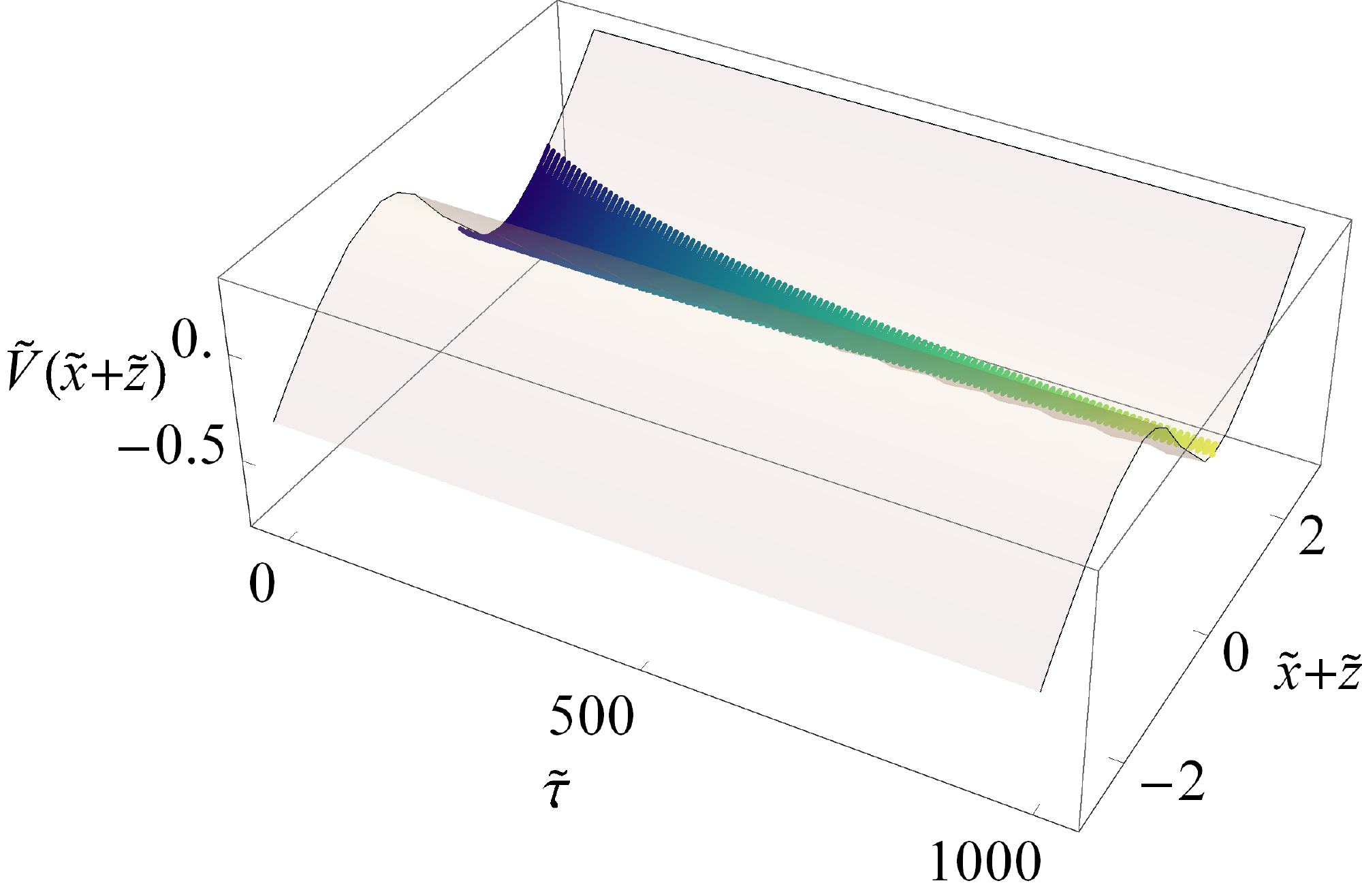}
  } \\
  \caption{\protect\subref{fig:PhotothermalStateSpace} The phase space with photothermal expansion and a negative photothermal expansion coefficient, taking parameters $(\tilde{g},\tilde{\Delta}_{\alpha},\tilde{\varepsilon})=(0.5,0,1/100)$ and $(\tilde{\zeta},\tilde{\gamma})=(-30,3\times 10^{-4})$. Time runs from blue to yellow; the trajectory begins at the bottom and exhibits decaying oscillations towards the steady state at the top. The dashed black line through the centre denotes the steady state for $\tilde{x}$, $\tilde{p}_x$ assuming a fixed value of $\tilde{z}$. \protect\subref{fig:PhotothermalPotential} The same trajectory, now plotted in the potential $\tilde{V}$.}\label{fig:PhotothermalTrajectories}
\end{figure}

Photothermal effects may also alter the stability of the equilibria. We discuss in \cref{sec:APhotothermal} that due to the small value of $\tilde{\gamma}$, these behave as a small perturbation on our system. We may still look at the mirror as moving in the potential $\tilde{V}(\tilde{x}+\tilde{z})$, only now as well as having a damping/anti-damping effect from the dynamics of the light field, we have an additional photothermal contribution. Considering a small oscillation around the potential-well minimum, the change in dimensionless energy of the system over one period is
\begin{equation}\label{eq:PhotothermalDamping}
  2\tilde{\rho}^2(1-\tilde{g})\tilde{g}^3\tilde{\zeta}\tilde{\gamma}+\mathcal{O}(\tilde{\gamma}^2),
\end{equation}
where $\tilde{\rho}$ is the size of the small perturbation from equilibrium. Thus positive photothermal expansion ($\tilde{\zeta}>0$) will provide an anti-damping effect, while negative photothermal expansion ($\tilde{\zeta}<0$) will give damping. This is similar to what is observed in conventional optomechanical systems, where engineering a photothermal effect with a negative sign in order to stabilise the system has also been proposed \cite{Kelley2015}. Note that \cref{eq:PhotothermalDamping} describes photothermal effects only, which will be in competition with the damping/anti-damping induced by the light field.

This result relied upon a number of approximations, and we can verify this with a numerical search of the unapproximated system. We compute the sign of the maximum eigenvalue around the steady state for various parameter regimes, and plot these \cref{fig:PhotothermalEigenvalues}. We find that $\tilde{x}^s_+$ is in general unstable for a positive $\tilde{\zeta}$, and stable for a negative value. This difference becomes more pronounced as $\tilde{\varepsilon}$ grows smaller, and the optical heating effect decreases. Interestingly, for a large positive photothermal expansion coefficient, decreasing $\tilde{\varepsilon}$ actually increases the anti-damping rate. 

As in \cref{fig:VisualisationPhaseSpace}, we can visualise the trajectories in phase space. Taking advantage of the almost two-dimensional nature of the dynamics we replace the $|\tilde{\alpha}|^2$-axis with the photothermal variable $\tilde{z}$, and plot this in \cref{fig:PhotothermalStateSpace}. In this case we consider a negative photothermal expansion coefficient, and observe decaying oscillations towards the equilibrium state. These are centered on the steady state for $\tilde{x}$, $\tilde{p}_x$ assuming a fixed value $\tilde{z}$. From \cref{eq:PTSteadyState} this is given by the line
\begin{equation}
  \tilde{x}=-\tilde{\Delta}_{\alpha}+\tilde{\zeta}\tilde{z}+\tilde{\sigma}.
\end{equation}
in the $\tilde{p}_x=0$ plane. Finally, we plot the trajectory in the dimensionless potential $\tilde{V}$ in \cref{fig:PhotothermalPotential}. In this case we need to consider the total displacement $\tilde{x}+\tilde{z}$, and with this we can see the photothermal damping stabilising the mirror oscillations towards the well minimum.

%}]}

%}]}

\section{Conclusion}%{[{
We have analysed a vertically oriented Fabry-P\'erot cavity where the upper mirror levitates on the intra-cavity radiation pressure force. Nondimensionalising the equations of motion, the behaviour was essentially determined by two dimensionless parameters: the effective gravity $\tilde{g}$ and the ratio $\tilde{\varepsilon}$ of the mechanical to optical timescales. Typically the optical timescale is much faster than the mechanical ($\tilde{\varepsilon}\ll 1$), and exploiting this we performed a perturbative expansion around $\tilde{\varepsilon}\approx 0$ to simplify the dynamics to the point where an analytic approach was tractable.

To zeroth order this gave an adiabatic approximation. We found an effective potential in which the upper mirror moved, and used this to calculate its frequencies of oscillation. This also gave us an intuition for the evolution of the dynamical variables in phase space. A first-order correction to the equations of motion allowed us to derive the effective damping or anti-damping rate caused by the dynamics of the cavity field. Comparing with simulations of the exact equations, our approximation was very accurate for $\tilde{\varepsilon}\sim 1/100$ and still qualitatively similar for $\tilde{\varepsilon}\sim 1/10$, which encompasses a wide range of realistic parameter regimes. 

From this analysis we concluded that the motion of the upper mirror would always be unstable if the cavity was driven by a single laser, with the rate at which oscillations were amplified linearly proportional to $\tilde{\varepsilon}$. We then explored using two input lasers with different amplitudes and detunings, and found all parameter regimes which allowed for stable trapping of the upper mirror. In these regions the confining potential well could be made relatively wide, but was always flat and shallow. We also analysed the effect of photothermal changes in the optical path length of the cavity. If heating of the mirror substrate lead to a decrease in the cavity length, working against radiation pressure force on the mirror, there was an anti-damping effect. If however photothermal effects increased the cavity length they could stabilise the system, even with only a single laser.

Stabilising the levitating mirror is crucial in order to use it as a metrological platform, and there are numerous possible ways that we could extend our analysis. One is to move beyond two lasers; it is shown in ref. \cite{Slatyer2016} that as you allow a greater number of input lasers to a conventional optomechanical system, you can approximate an arbitrarily-shaped potential. It is less clear however this can be done for our system while also ensuring the potential is damping, and it is likely that any multi-laser potential wells which are both trapping and damping can be made very wide, but will maintain their shallow and flat nature. A promising avenue would be feedback control of the system, which should be able to stabilise it even with only a single laser. Once stabilisation has been achieved, a detailed study can be performed on the various ways the levitating mirror could be used for sensing, and how to best utilise the many unique features of this platform.

%}]}

\section{Acknowledgements}%{[{
This research was funded by the Australian Research Council Centre of Excellence for Quantum Computation and Communication Technology (grant no. CE110001027). PKL acknowledges support from the ARC Laureate Fellowship FL150100019, and RL is supported by an Australian Government Research Training Program (RTP) scholarship. RL would also like to thank Craig M Savage for introducing him to this project, and for all that Craig has done over the years.

%}]}

%}]}

%{[{ Appendix
\appendix 

\section{Derivation of one-dimensional non-dimensionalised equations}\label{sec:ANondimensionalisation}%{[{
Here we describe the nondimensionalisation from \cref{sec:MainNondimensionalisation} in more detail. We will introduce a natural length scale $\ell$, frequency $\nu$, and cavity amplitude $A$, defined as
\begin{equation}\label{eq:AIdealNDScales}
  \ell=\frac{\delta\omega/2}{G},\;\nu^2=\frac{\hbar GA^2}{m\ell},\;A=\frac{1}{\delta\omega/2}\sqrt{\frac{\kappa_iP_{\alpha}}{\hbar\omega_0}}.
\end{equation}
To understand these scales, we solve for the steady state of $\alpha$ from \cref{eq:BaseEquations} giving the Lorentzian profile (setting $\Delta_{\alpha}=0$ for simplicity):
\begin{equation}
  |\alpha|^2=\frac{(\delta\omega/2)^2A^2}{(\delta\omega/2)^2+(Gx)^2}=\frac{A^2}{1+\left(\frac{Gx}{\delta\omega/2}\right)^2}.
\end{equation}
We thus see that $A$ is the maximum intra-cavity amplitude. Moreover, at $x=(\delta\omega/2)/G=\ell$ the amplitude-squared has dropped to half of its maximum value, or equivalently $\ell$ is the distance by which the mirror moves to create a cavity detuning of one half-linewidth. In a classical optomechanical system the optical spring constant is given by the first derivative of the radiation pressure force \cite[\S V.A]{Aspelmeyer2014}. Its value depends on $x$, and at $x=\ell$ we will have an optical spring constant
\begin{equation}
  \begin{aligned}
    \kappa_{os}(\ell) &= \frac{\partial}{\partial x}\left(-\hbar G|\alpha|^2\right), \\
      &=  \frac{\hbar G^2A^2}{\delta\omega}=  2m\frac{\hbar GA^2}{m\ell}, \\ &=  m\left(\frac{\nu}{\sqrt{2}}\right)^2.
  \end{aligned}
\end{equation}
Comparing this with the usual relation $\kappa=m\omega^2$ for a harmonic oscillator with spring constant $\kappa$, mass $m$, and natural frequency $\omega$, we see that $\nu$ is $1/\sqrt{2}$ multiplied by the optical spring frequency at one half-linewidth's displacement. The `mechanical timescale' $\nu$ gives timescale of the upper mirror's oscillations, while the `optical timescale' $\delta\omega/2$ describes the speed of the cavity field's dynamics.

The dimensionless dynamical variables (denoted by a tilde) are:
\begin{equation}\label{eq:AIdealNDSubstitutions}
  \tilde{\tau}=\nu t,\; \tilde{x}=\frac{x}{\ell},\; \tilde{p}=\frac{p}{m\ell\nu},\;\tilde{\alpha}=\frac{\alpha}{A}.
\end{equation}
For the cavity field we will have $0<|\tilde{\alpha}|^2<1$. Furthermore during levitation the mirror variables $\tilde{x}$ and $\tilde{p}_x$ will typically be of order $1$, as if $\tilde{x}\gg 1$ the cavity detuning will be far from resonance so the optical force will approach zero.

We can now write the equations of motion in terms of the dimensionless parameters, letting a prime denote the derivative with respect to dimensionless time $\tilde{\tau}$:
\begin{equation}
  \begin{aligned}
    \ell\nu\tilde{x}' &= \frac{p_x}{m}, \\
    \tilde{x}' &= \tilde{p}_x.
  \end{aligned}
\end{equation}
\begin{equation}\label{eq:ANonDimingPx}
  \begin{aligned}
    m\ell\nu^2\tilde{p}_x' &= -mg+\hbar GA^2|\tilde{\alpha}|^2, \\
    \tilde{p}_x' &= -\frac{g}{\ell\nu^2}+\frac{\hbar GA^2}{m\ell\nu^2}|\tilde{\alpha}|^2, \\
      &= -\tilde{g}+|\tilde{\alpha}|^2.
  \end{aligned}
\end{equation}
\begin{equation}
  \begin{aligned}
    A\nu\tilde{\alpha}' &= \left(-\frac{\delta\omega}{2}+i\left(\Delta_{\alpha}+G\ell\tilde{x}\right)\right)A\tilde{\alpha}+\sqrt{\frac{\kappa_iP_{\alpha}}{\hbar\omega_0}}, \\
    \tilde{\alpha}' &= \frac{\delta\omega/2}{\nu}\left[i\left(\tilde{\Delta}_{\alpha}+\tilde{x}\right)\tilde{\alpha}-\tilde{\alpha}+\frac{1}{A(\delta\omega/2)}\sqrt{\frac{\kappa_iP_{\alpha}}{\hbar\omega_0}}\right], \\
    &= \left[i\left(\tilde{\Delta}_{\alpha}+\tilde{x}\right)\tilde{\alpha}-\tilde{\alpha}+1\right]/\tilde{\varepsilon}.
  \end{aligned}
\end{equation}

We have defined three dimensionless parameters:
\begin{equation}
  \tilde{g}=\frac{g}{\ell\nu^2},\;\tilde{\varepsilon}=\frac{\nu}{\delta\omega/2},\;\tilde{\Delta}_{\alpha}=\frac{\Delta_{\alpha}}{\delta\omega/2}.
\end{equation}
The meanings of these are explored in the main text of \cref{sec:MainNondimensionalisation}.

To summarise, the dimensionless equations of motion are
\begin{equation}\label{eq:ANDBaseEquations}
  \begin{aligned}
    \tilde{x}'      &= \tilde{p}_x, \\
    \tilde{p}_x'    &= -\tilde{g}+|\tilde{\alpha}|^2, \\
    \tilde{\alpha}' &= \left[i(\tilde{\Delta}_{\alpha}+\tilde{x})\tilde{\alpha}-\tilde{\alpha}+1\right]/\tilde{\varepsilon},
  \end{aligned}
\end{equation}
where $0<\tilde{g},|\tilde{\alpha}|^2<1$, and typically $\tilde{\varepsilon}\ll1$. Moreover if $\tilde{\Delta}_{\alpha}$ is constant, we can eliminate it from our dynamics by defining $\tilde{\chi}=\tilde{\Delta}_{\alpha}+\tilde{x}$.

%}]}

\section{Expanding the equations of motion in orders of $\tilde{\varepsilon}$}\label{sec:AExpandingEpsilon}%{[{
The parameter $\tilde{\varepsilon}$ will typically be quite small, corresponding to the optical field evolving much faster than the mechanical motion of the mirror. Our dynamics \cref{eq:NDBaseEquations} thus have a separation of timescales, also known in literature as a `slow-fast system'. We can exploit this to simplify the dynamics in the manner of \cite[\S 2]{VanKampen1985}, by expanding in a power series of $\tilde{\varepsilon}$.

We begin with
\begin{equation}
  \tilde{\alpha}=\tilde{\alpha}_0+\tilde{\varepsilon}\tilde{\alpha}_1+\mathcal{O}(\tilde{\varepsilon}^2),
\end{equation}
and substitute this into the equation of motion for $\tilde{\alpha}'$ from \cref{eq:ANDBaseEquations}:
\begin{equation}
  \begin{aligned}
    &\tilde{\alpha}_0'+\tilde{\varepsilon}\tilde{\alpha}_1'+\mathcal{O}(\tilde{\varepsilon}^{2})\\
    &\;\;=\tilde{\varepsilon}^{-1}\left(i(\tilde{\Delta}_{\alpha}+\tilde{x})-1\right)\left(\tilde{\alpha}_0+\tilde{\varepsilon}\tilde{\alpha}_1+\mathcal{O}(\tilde{\varepsilon}^{2})\right)+\tilde{\varepsilon}^{-1}.
  \end{aligned}
\end{equation}
Equating terms of order $\tilde{\varepsilon}^{-1}$ (or equivalently multiplying both sides by $\tilde{\varepsilon}$ then taking the limit $\tilde{\varepsilon}\rightarrow 0$) gives
\begin{equation}
  0 = \left(i(\tilde{\Delta}_{\alpha}+\tilde{x})-1\right)\tilde{\alpha}_0+1,
\end{equation}
and solving this for $\tilde{\alpha}_0$ recovers the adiabatic approximation we discussed in \cref{sec:MainAdiabatic}:
\begin{equation}\label{eq:APowerSeriesAdiabatic}
  \tilde{\alpha}_0=\frac{1}{1-i(\tilde{\Delta}_{\alpha}+\tilde{x})}.
\end{equation}
We can thus see that the adiabatic equations of motion \cref{eq:AdiabaticIdealEOM} correspond to expanding the full equations of motion \cref{eq:ANDBaseEquations} to zeroth order in $\tilde{\varepsilon}$.

The terms of order $1$ give
\begin{equation}
  \tilde{\alpha}_0' = \left(i(\tilde{\Delta}_{\alpha}+\tilde{x})-1\right)\tilde{\alpha}_1, \\
\end{equation}
which has solution
\begin{equation}\label{eq:APowerSeriesA1Primitive}
  \tilde{\alpha}_1=\frac{-\tilde{\alpha}_0'}{1-i(\tilde{\Delta}_{\alpha}+\tilde{x})}.
\end{equation}
Differentiating \cref{eq:APowerSeriesAdiabatic} using $\tilde{\Delta}_{\alpha}'=\tilde{s}$ and $\tilde{x}'=\tilde{p}_x$ gives
\begin{equation}
  \tilde{\alpha}_0'=\frac{i\left(\tilde{s}+\tilde{p}_x\right)}{\left(1-i(\tilde{\Delta}_{\alpha}+\tilde{x})\right)^2}.
\end{equation}
Substituting this into \cref{eq:APowerSeriesA1Primitive} then gives us $\tilde{\alpha}_1$, which we can use to compute $|\tilde{\alpha}|^2$ to first-order in $\tilde{\varepsilon}$:
\begin{equation}\label{eq:AAlpha2FirstOrder}
  \left\lvert\tilde{\alpha}^2\right\rvert=\frac{1}{1+\left(\tilde{x}+\tilde{\Delta}_{\alpha}\right)^2}+\tilde{\varepsilon}\frac{4(\tilde{p}_x+\tilde{s})(\tilde{x}+\tilde{\Delta}_{\alpha})}{\left(1+(\tilde{x}+\tilde{\Delta}_{\alpha})^2\right)^3}+\mathcal{O}\left(\tilde{\varepsilon}^{2}\right).
\end{equation}
The equations of motion to first-order in $\tilde{\varepsilon}$ are then
\begin{equation}\label{eq:AFirstOrderEOM}
  \begin{aligned}
    \tilde{x}'    &= \tilde{p}_x, \\
    \tilde{p}_x'  &= -\tilde{g}+\frac{1}{1+\left(\tilde{x}+\tilde{\Delta}_{\alpha}\right)^2}+\tilde{\varepsilon}\frac{4(\tilde{p}_x+\tilde{s})(\tilde{x}+\tilde{\Delta}_{\alpha})}{\left(1+(\tilde{x}+\tilde{\Delta}_{\alpha})^2\right)^3}.
  \end{aligned}
\end{equation}

%}]}

\section{The adiabatic approximation}\label{sec:AAdiabatic} %{[{
Here we provide more detail on the results from the adiabatic approximation in \cref{sec:MainAdiabatic}. We begin with the equations of motion \cref{eq:AdiabaticIdealEOM} 
\begin{equation}
  \begin{aligned}
    \tilde{x}'    &= \tilde{p}_x, \\
    \tilde{p}_x'  &= -\tilde{g}+\frac{1}{1+(\tilde{\Delta}_{\alpha}+\tilde{x})^2}.
  \end{aligned}
\end{equation}
Partially integrating the equation of motion for $\tilde{p}_x'$ with respect to $\tilde{x}$ gives
\begin{equation}\label{eqA:AdiabaticPotential}
  -\int\left(\frac{d\tilde{p}_x}{d\tilde{\tau}}\right)d\tilde{x}=\tilde{g}\tilde{x}-\arctan\left(\tilde{\Delta}_{\alpha}+\tilde{x}\right).
\end{equation}
As the dynamics are unchanged by a constant shift in the potential, we define the effective dimensionless potential energy as
\begin{equation}\label{eqA:DimPotential}
  \tilde{V}(\tilde{x})=\tilde{g}(\tilde{\Delta}_{\alpha}+\tilde{x})-\arctan\left(\tilde{\Delta}_{\alpha}+\tilde{x}\right).
\end{equation}
Then we can verify that $\tilde{p}_x'=-\nabla \tilde{V}(\tilde{x})$. Adding a `kinetic energy' term $(\tilde{p}_x')^2/2$, the total dimensionless energy is defined as
\begin{equation}\label{eqA:AdiabaticEnergy}
  \tilde{\mathcal{E}}(\tilde{x},\tilde{p}_x)=\frac{\tilde{p}_x^2}{2}+\tilde{V}(\tilde{x}).
\end{equation}
We can verify that $\tilde{\mathcal{E}}$ is conserved by directly computing:
\begin{equation}
  \begin{aligned}
    \frac{d\tilde{\mathcal{E}}}{d\tilde{\tau}}&=\tilde{p}_x\frac{d\tilde{p}_x}{d\tilde{\tau}}+\frac{d\tilde{V}}{d\tilde{x}}\frac{d\tilde{x}}{d\tilde{\tau}}, \\
      &=\tilde{p}_x\left(-\frac{d\tilde{V}(\tilde{x})}{d\tilde{x}}\right)+\frac{d\tilde{V}}{d\tilde{x}}\left(\tilde{p}_x\right), \\
      &=0.
  \end{aligned}
\end{equation}
Thus in the adiabatic limit, the levitated mirror will behave like a classical particle moving in the potential \cref{eqA:DimPotential}.

To find the frequency of these oscillations we follow the procedure outlined in \cite[\S 2]{Amore}. Suppose the mirror oscillates between two points $\tilde{x}_M$ and $\tilde{x}_P$. We necessarily have $\tilde{V}(\tilde{x}_M)=\tilde{V}(\tilde{x}_P)$, and $-\tilde{\sigma}<\tilde{x}_M<\tilde{\sigma}<\tilde{x}_P$. The dimensionless period $\tilde{\mathcal{T}}$ of an oscillation is:
\begin{equation}
  \tilde{\mathcal{T}}=\oint d\tilde{\tau}=2\int_{\tilde{x}_M}^{\tilde{x}_P}\left(\frac{d\tilde{\tau}}{d\tilde{x}}\right)d\tilde{x},
\end{equation}
which can be re-dimensionalised as $\mathcal{T}=\tilde{\mathcal{T}}/\nu$. We have $\frac{d\tilde{\tau}}{d\tilde{x}}=\left(\frac{d\tilde{x}}{d\tau}\right)^{-1}=(\tilde{p}_x)^{-1}$, and we can write this in terms of the dimensionless energy using \cref{eq:AdiabaticEnergy}:
\begin{equation}\label{eq:AdiaPeriod}
  \tilde{\mathcal{T}}=\sqrt{2}\int_{\tilde{x}_M}^{\tilde{x}_P}\frac{d\tilde{x}}{\sqrt{\tilde{\mathcal{E}}(\tilde{x},\tilde{p}_{x})-\tilde{V}(\tilde{x})}}.
\end{equation}
Since energy is conserved in the adiabatic regime, over the whole period we have
\begin{equation}
  \tilde{\mathcal{E}}(\tilde{x},\tilde{p}_x)=\tilde{V}(\tilde{x}_M),
\end{equation}
and so the integral in \cref{eq:AdiaPeriod} can be evaluated. To find all periods, we choose $\tilde{x}_M\in(-\tilde{\sigma},\tilde{\sigma})$, and then numerically solve the transcendental equation $\tilde{V}(\tilde{x})=\tilde{V}(\tilde{x}_M)$ whose solution is $\tilde{x}_P$ (choosing the root which lies to the right of $\tilde{\sigma}$). Once $\tilde{x}_M$ and $\tilde{x}_P$ are in hand we may numerically compute the integral in \cref{eq:AdiaPeriod}. The range of possible amplitudes increases as $\tilde{g}$ decreases, since the width of the potential well increases. We plot the frequencies in \cref{fig:OscillationFrequency} (rather than the period, since $\tilde{\mathcal{T}}$ approaches infinity for the maximum amplitude oscillations). Note that the frequency of oscillation depends solely on the dimensionless parameter $\tilde{g}$ and the amplitude of oscillation, and furthermore the frequency decreases as the amplitude increases.

So far we have assumed that the input laser detuning $\tilde{\Delta}_{\alpha}$ is static. We will now briefly investigate  allowing $\tilde{\Delta}_{\alpha}$ to vary linearly with $\tilde{\tau}$, which can occur when we scan the cavity length or laser frequency:
\begin{equation}
  \tilde{\Delta}_{\alpha}(\tilde{\tau})=\tilde{\Delta}_{\alpha0}+\tilde{s}\tilde{\tau},
\end{equation}
where $\tilde{\Delta}_{\alpha0}$ is the initial detuning and $\tilde{s}$ the scan speed. Such a sweeping is commonly used to probe and characterise the optical system, and will correspond to shifting the potential shown in \cref{fig:DimPotential} horizontally at the speed $\tilde{s}$. 

First suppose that $\tilde{\Delta}_{\alpha0}$ is negative (i.e. the laser is red-detuned compared to the cavity), and $\tilde{s}$ is positive. The mirror begins at $\tilde{x}=0$, and we assume this to be sitting on a stand which prevents it from falling below this point. The potential $\tilde{V}$ is initially centred at $-\tilde{\Delta}_{\alpha 0}>0$, and moves to the left at speed $\tilde{s}$. The force due to the potential on the mirror is initially downwards, so the mirror will be stationary due to the stand. When $\tilde{\Delta}_{\alpha}(\tilde{\tau})=-\tilde{\sigma}$ the local maximum of the potential will pass the mirror leading to a net upwards force, and so it will begin to oscillate on the stand. Eventually however the potential will keep moving and the force on the mirror will become negative; the oscillations will cease and it will rest on the stand.

The other case is for $\tilde{\Delta}_{\alpha0}>0$ and $\tilde{s}<0$, where the centre of the potential begins to the left of the mirror and moves towards the right. If the scan speed is slow enough, the mirror will be `picked up' by the potential well and there will be oscillations centred about $\tilde{x}=\tilde{\Delta}_{\alpha}(\tilde{\tau})$. If the scan is too fast however the moving potential well will not be able to pick up the mirror, so after a brief period of oscillation the force will become downwards again and oscillations will cease. The critical scan velocity $\tilde{s}_{c}$ below which the mirror is picked up can be found by moving to a reference frame where the potential is stationary, while the mirror moves to the left at speed $\tilde{s}$. Because of the stand, the mirror doesn't begin moving until the force is in the upwards direction, and so the initial mirror position in this frame must be at the local minimum of the potential well. In this case for the mirror to be trapped by the potential we require its kinetic energy to be less than the potential barrier between $\tilde{\sigma}$ and $-\tilde{\sigma}$:
\begin{equation}
  \frac{\tilde{s}^2}{2}\le 2\left(\arctan(\tilde{\sigma})-\tilde{\sigma}\tilde{g}\right),
\end{equation}
which gives
\begin{equation}
  \tilde{s}_c=2\sqrt{\arctan(\tilde{\sigma})-\tilde{\sigma}\tilde{g}}.
\end{equation}

%}]}

\section{Calculation of the photothermal dissipation rate}\label{sec:APhotothermal}%{[{
Here we will investigate whether photothermal effects lead to damping or anti-damping. To focus on the photothermal effect alone, we eliminate the influence of the optical field by making an adiabatic approximation. This is justified as typically the dynamics of the light field, described by $\tilde{\varepsilon}^{-1}$, are fast compared to the photothermal relaxation rate $\tilde{\gamma}$. With this we have equations of motion
\begin{equation}\label{eq:APhotothermalOnlyEOM}
  \begin{aligned}
    \tilde{x}'        &= \tilde{p}_x, \\
    \tilde{p}_x'      &= -\tilde{g}+\frac{1}{1+(\tilde{\Delta}_{\alpha}+\tilde{x}+\tilde{z})^2}, \\
    \tilde{z}'        &= -\tilde{\gamma}\left[\tilde{z}+\tilde{\zeta}\frac{1}{1+(\tilde{\Delta}_{\alpha}+\tilde{x}+\tilde{z})^2}\right].
  \end{aligned}
\end{equation}
We then linearise the system \cref{eq:APhotothermalOnlyEOM} about the steady state $\tilde{x}^s_+$, leading to equations of motion
\begin{equation}\label{eq:ALinearisedPhotothermal}
  \begin{aligned}
    \delta\tilde{x}'    &= \delta\tilde{p}_x, \\
    \delta\tilde{p}_x'  &= -\tilde{\omega}^2(\delta\tilde{x}+\delta\tilde{z}), \\
    \delta\tilde{z}'    &= \tilde{\gamma}\tilde{\zeta}\tilde{\omega}^2\delta\tilde{x}-\tilde{\lambda}\delta\tilde{z},
  \end{aligned}
\end{equation}
with constants defined as
\begin{equation}
  \tilde{\omega}^2=2\tilde{g}^2\sqrt{\frac{1}{\tilde{g}}-1},\;\tilde{\lambda}=\tilde{\gamma}(1-\tilde{\omega}^2\tilde{\zeta}).
\end{equation}
While the coupled linear system \cref{eq:ALinearisedPhotothermal} can be solved directly, this results in a very complicated expression. We can however derive a simpler approximate solution, taking advantage of the small value of $\tilde{\gamma}$.

Suppose that the mirror is perturbed slightly from $\tilde{x}^s_+$. From our analysis of the adiabatic system in  \cref{sec:MainAdiabatic} we know that the mirror will oscillate around the steady state. As this happens, this will induce oscillations in the intra-cavity field, and thus photothermal expansion $\delta\tilde{z}$. Due to the small value of $\tilde{\gamma}$, the oscillations of $\delta\tilde{z}$ will be much smaller than those of $\delta\tilde{x}$ and $\delta\tilde{p}_x$. A zeroth order approximation will thus be to take $\delta\tilde{z}\approx\mathcal{O}(\tilde{\gamma})\approx 0$, in which case we find from \cref{eq:ALinearisedPhotothermal}
\begin{equation}\label{eq:APTZerothOrder}
  \begin{aligned}
    \delta\tilde{x}(\tilde{\tau})   &= \tilde{\rho}\sin(\tilde{\omega}\tilde{\tau}), \\
    \delta\tilde{p}_x(\tilde{\tau}) &= \tilde{\rho}\tilde{\omega}\cos(\tilde{\omega}\tilde{\tau}), \\
    \delta\tilde{z}(\tilde{\tau})   &= 0,
  \end{aligned}
\end{equation}
where $\tilde{\rho}$ is the size of the perturbation to $\delta\tilde{x}$. We can then substitute \cref{eq:APTZerothOrder} into the equations of motion \cref{eq:ALinearisedPhotothermal} to calculate the error, which we find to be $\mathcal{O}(\tilde{\gamma})$.

Next we use \cref{eq:APTZerothOrder} to generate the first-order approximation. We assume $\delta\tilde{x}$, $\delta\tilde{p}_x$ take the form given in \cref{eq:APTZerothOrder}, and then solve the equation of motion for $\delta\tilde{z}'$ which yields
\begin{equation}\label{eq:APTZFirstOrder}
  \delta\tilde{z}(\tilde{\tau}) = \frac{\tilde{\rho}\tilde{\omega}^2\tilde{\gamma}\tilde{\zeta}}{\tilde{\omega}^2+\tilde{\lambda}^2}\left(\tilde{\lambda}\sin(\tilde{\omega}\tilde{\tau})-\tilde{\omega}\cos(\tilde{\omega}\tilde{\tau})\right).
\end{equation}
We then substitute in this form for $\delta\tilde{z}$ into the equations of motion \cref{eq:ALinearisedPhotothermal}, and derive
\begin{equation}\label{eq:APTXPFirstOrder}
  \begin{aligned}
    \delta\tilde{x}(\tilde{\tau}) &= \tilde{\rho}\sin(\tilde{\omega}\tilde{\tau}) \\
      &\phantom{=}+\tilde{\rho}\tilde{\gamma}\tilde{E}\left(\tilde{C}_+(\tilde{\tau})\cos(\tilde{\omega}\tilde{\tau})-\tilde{D}_-(\tilde{\tau})\sin(\tilde{\omega}\tilde{\tau})\right), \\
    \delta\tilde{p}_x(\tilde{\tau}) &= \tilde{\rho}\omega\cos(\tilde{\omega}\tilde{\tau}) \\
      &\phantom{=}+\tilde{\rho}\tilde{\gamma}\tilde{\omega}\tilde{E}\left(\tilde{D}_+(\tilde{\tau})\cos(\tilde{\omega}\tilde{\tau})+\tilde{C}_-(\tilde{\tau})\sin(\tilde{\omega}\tilde{\tau})\right), \\
  \end{aligned}
\end{equation}
where
\begin{equation}
  \begin{aligned}
    \tilde{E} &= \frac{\tilde{\zeta}\tilde{\omega}^2}{4(\tilde{\omega}^2+\tilde{\lambda}^2)}, \\
    \tilde{C}_{\pm}(\tilde{\tau}) &= \tilde{\omega}(1\pm 2\tilde{\lambda}\tilde{\tau}), \\
    \tilde{D}_{\pm}(\tilde{\tau}) &= \tilde{\lambda}\pm 2\tilde{\omega}^2\tilde{\tau}.
  \end{aligned}
\end{equation}
Substituting \cref{eq:APTZFirstOrder} and \cref{eq:APTXPFirstOrder} into the equations of motion \cref{eq:ALinearisedPhotothermal} the error is found to be of order $\mathcal{O}(\tilde{\gamma}^2\tilde{\tau})$, which is negligible for small times. We can thus use these approximate solutions to see how the photothermal expansion will affect the stability of $\tilde{x}^s_+$.

Accounting for photothermal expansion, the energy of the mirror is now
\begin{equation}
  \tilde{\mathcal{E}}=\frac{\tilde{p}_x^2}{2}+\tilde{g}\tilde{x}-\tan^{-1}(\tilde{x}+\tilde{z}).
\end{equation}
Assuming we are in the neighbourhood of the steady state $(\tilde{x}^s_+,\tilde{p}_x^s,\tilde{z}^s)$, we have to second order
\begin{equation}
  \tilde{\mathcal{E}}=\tilde{\mathcal{E}}^s-\tilde{g}\delta\tilde{z}+\frac{\delta\tilde{p}_x^2}{2}+\frac{\tilde{\omega}^2}{2}(\delta\tilde{x}+\delta\tilde{z})^2+\mathcal{O}(\delta^3),
\end{equation}
where $\tilde{\mathcal{E}}^s$ is the steady state energy. If the mirror position is perturbed a distance $\tilde{\rho}$ from equilibrium, it's motion will approximate periodic oscillation with period 
\begin{equation}
  \tilde{T}=\frac{2\pi}{\tilde{\omega}}.
\end{equation}
The average heating over a single oscillation can then be found via
\begin{equation}
  \frac{1}{\tilde{T}}\int_0^{\tilde{T}}\frac{\mathrm{d}\mathcal{\tilde{E}}}{\mathrm{d}\tilde{\tau}}\mathrm{d}\tilde{\tau}.
\end{equation}
Using the derived forms \cref{eq:APTZFirstOrder} and \cref{eq:APTXPFirstOrder}, we find this to be
\begin{equation}
  \frac{\tilde{\omega}^4}{2}\tilde{\rho}^2\tilde{\zeta}\tilde{\gamma}+\mathcal{O}(\tilde{\gamma}^2)=2(1-\tilde{g})\tilde{g}^3\tilde{\rho}^2\tilde{\zeta}\tilde{\gamma}+\mathcal{O}(\tilde{\gamma}^2).
\end{equation}
Thus photothermal expansion induces anti-damping at a rate linearly proportional to $\tilde{\zeta}$: positive photothermal expansion will provide an anti-damping effect, while negative photothermal expansion will give damping.

%}]}

%}}}

\bibliography{bibliography}

\end{document}